\documentclass[
aip,
physrev,
amsmath,
amssymb,
reprint,
]{revtex4-1}

\usepackage{graphicx}
\usepackage{dcolumn}
\usepackage{bm}
\usepackage[utf8]{inputenc}
\usepackage[T1]{fontenc}
\usepackage{mathptmx}

\usepackage{paralist}
\usepackage{xcolor}
\usepackage{bbold}
\usepackage[inline]{enumitem}

\usepackage{physics}
\usepackage{braket}
\usepackage{bm}
\usepackage{etoolbox}

\newcommand{\eq}{Eq.~}
\newcommand{\fig}{Fig.~}

\newcommand{\iu}{\mathrm{i}}
\newcommand{\D}{\,\mathrm{d}}

\newcommand{\up}{\uparrow}
\newcommand{\down}{\downarrow}
\newcommand{\hamiltonian}{Hamiltonian}
\newcommand{\iw}{\iu\omega_n}

\newcommand{\imp}{\mathrm{imp}}
\newcommand{\wdmft}{$\omega$-DMFT}
\newcommand{\weiss}{\mathrm{Weiss}}

\newcommand{\pimp}{P_\mathrm{imp}}

\begin{document}
\title{Frequency-dependent and algebraic bath states for a Dynamical Mean-Field Theory with compact support}
\author{Max~Nusspickel}
\author{George~H.~Booth}
\email{george.booth@kcl.ac.uk}
\affiliation{
Department of Physics,
King's College London,
Strand, London, WC2R 2LS, U.K.}

\date{\today}

\begin{abstract}

We demonstrate an algebraic construction of frequency-dependent bath orbitals
which can be used in a robust and rigorously self-consistent 
DMFT-like embedding method, here called \wdmft{}, suitable for use with Hamiltonian-based impurity solvers.
These bath orbitals are designed to exactly reproduce the hybridization of the
impurity to its environment, while allowing for a systematic expansion of this bath
space as impurity interactions couple frequency points.
%
%
In this way, the difficult non-linear fit of bath parameters
necessary for many Hamiltonian-formulation impurity solvers in DMFT is avoided, while
the introduction of frequency dependence in this bath space is shown to
allow for more compact bath sizes.
This has significant potential use with a number of new, emerging Hamiltonian solvers which allow 
for the embedding of large impurity spaces within a DMFT framework.
We present results of the \wdmft{} approach for the Hubbard model
on the Bethe lattice, a 1D chain, and the 2D square lattice,
which show excellent agreement with standard DMFT results, 
with fewer bath orbitals and more compact support for the hybridization representation 
in the key impurity model of the method.

%
%
%
%
%
%
%
%
%
\end{abstract}

\maketitle

\section{Introduction} \label{sec:intro}
Strongly correlated electron systems exhibit some of the most
interesting macroscopic manifestations of quantum collective behavior,
and as such have been of major interest both in the
condensed matter physics and quantum chemistry community\cite{StronglyCorrMat}.
Often these correlation effects manifest from strong interactions of electrons
within local atomic-like degrees of freedom which lie close to the Fermi energy.
Describing the strong interactions within these local `impurity' orbitals while simultaneously describing their
delocalized character into the wider system with which they are hybridized, is the driving
principle of many quantum embedding methods.
The dynamical mean-field theory (DMFT) has been successful
in this regard, initially applied to the Hubbard model on an infinitely coordinated
lattice, where its approximation of a strictly local self-energy is formally exact
\cite{Metzner1989,Zhang1993,Georges1996}.
%
%
%
The core of the DMFT framework involves an impurity solver, where the dynamics of the
interacting impurity is computed in the presence of some representation of its coupling
to its environment via the hybridization function. A common approach is to rely on
continuous-time quantum Monte-Carlo (CT-QMC) techniques, which allow for an integration over
the entire action of this hybridization in a stochastic technique.\cite{Gull2011}
Alternatively, there
exists a growing body of research focussing on approaches to solve the impurity problem
which exists within a Hamiltonian formulation. In order to do this, the
hybridization must be discretized as a finite set of bath orbitals which are explicitly
coupled to the impurity space. Traditionally, this resulting problem was solved with exact
diagonalization (ED) methods (known as full configuration-interaction in the quantum chemistry
community).
This has some advantages over CT-QMC, including the ability to solve the problem at strictly
zero electronic temperature, the ability to compute real-frequency dynamical quantities without
relying on analytic continuation, the absence of stochastic errors, and the ability to seamlessly
deal with arbitrary forms of the two-body impurity interactions. However, the price to pay is the
discretization of the hybridization in terms of these bath orbitals, and the exponential
cost of the solver as the number of bath orbitals is increased, necessitating a very compact number
of orbitals to represent this continuous function.
This places severe limitations onto cluster DMFT methods, since one has to
compromise between the description of long-range correlation by enlarging
the impurity size and a faithful representation of the hybridization by including
more bath orbitals, which can often limit the true physics of the system from emerging.
To mitigate this, a number of lower cost Hamiltonian-formulation solvers are now being developed,
which allow for a larger number of bath orbitals to be included, while admitting minimal
error into the description of the impurity dynamics. These are based on selected configuration interaction,
density matrix renormalization group, stochastic methods, and quantum chemical methodologies among
others\cite{Lu2017,Granath2012,Chan2012,Millis17,ACISolver,Wolf15,MPS15,MPS15_2,ChanCC,ZgidCC,Rhodes19,KPFCIQMC,Booth12,FCIQMC_Realtime}.
However, while these approaches can allow more bath orbitals, the difficulty in finding an appropriate and compact
bath space, especially in situations where there are a larger number of impurities or a complex
hybridization with substantial non-diagonal contributions, remains significant. This is due to the non-convex numerical optimization
which is required, which can frequently result in local minima when constrained to a small number of bath orbitals, and 
whose many solutions when matching on the Matsubara axis can make a consistent level of description hard to achieve via this approach.

In this work we circumvent these problems via an algebraic approach to generate a compact set of explicit bath orbitals
which define an impurity auxiliary system with well-defined properties, exactly capturing the hybridization and ensuring a
rapid and systematic convergence of the bath space compared to traditional ED-DMFT. In addition, the approach avoids the non-linear
numerical fit of bath orbitals of the impurity problem. The price we pay for these desirable properties is that the constructed bath space
changes with frequency and therefore an impurity model has to be solved for each frequency point of interest. 
However, this property of the method rationalizes the claim of being able to capture the exact hybridization despite a finite bath space,
as well as the prospect of a more compact bath space at each frequency point and reduction in the bath discretization error. The method is also exact in the non-interacting limit, with the cluster and lattice
Green's function identical by construction due to the exact hybridization of the bath space.
Due to the frequency-dependence of the auxiliary impurity problems, we term this approach \wdmft{}.

The method bears relation to a number of approaches in the literature that is worth expanding upon. The density-matrix
embedding theory (DMET) takes a different approach to the embedding problem, which has also had a number of notable successes.\cite{Knizia2012,PhysRevB.93.035126,Zheng1155}
In this method, the impurity problem has to be solved for the one-particle reduced density matrix only,
reducing the computational cost greatly in comparison to DMFT. The coupling to the environment in order to reproduce
the projection of this lattice density matrix is also algebraic and compact, further reducing the cost. This is obtained 
from the Schmidt decomposition of a mean-field wavefunction, ensuring an exact
matching of the uncorrelated density-matrix between impurity model and lattice.
However, the absence of dynamical effects restricts the local physics which can be
described in DMET compared to DMFT.
For instance, the lack of a self-consistent and dynamical self-energy
in the lattice precludes the description of correlated physics such as local fluctuating moments,
e.g. Mott insulating phases in the lattice, which are
instead mimicked via symmetry-breaking. In order to describe dynamical correlation functions, it was shown in Ref.~\onlinecite{Booth2015} that the DMET procedure could be augmented with a
single-shot, non-self-consistent calculation of the Green's function, using bath orbitals derived from a Schmidt decomposition of a mean-field response function. However, the lack of dynamical effects in the lattice again precluded a self-consistent treatment. 

In order to ameliorate these limitations of DMET, recently the `energy-weighted density-matrix embedding theory' (EwDMET) was proposed as a way to add a controllable resolution of dynamical fluctuations between the impurity and environment, while still retaining a computationally efficient, static formulation.\cite{Fertitta2018,Fertitta2019}
In this approach local energy-weighted density matrices are matched between the
impurity and lattice, providing a limited,
yet systematically improvable dynamical resolution of the impurity propagator.
Furthermore, correlation effects in the lattice are self-consistently described by an
effective dynamical self-energy, represented via additional auxiliary degrees of freedom
on the lattice.
True correlation-driven Mott physics was therefore demonstrated in a DMET context,
even for a single impurity-site, without the necessity of symmetry-breaking.

The \wdmft{} approach detailed in this work can be considered an extension of the EwDMET method for spectral embedding, allowing for full dynamical effects of the impurity, and building on and generalizing the previous work of the construction of (non-self-consistent) bath spaces for spectral DMET\cite{Booth2015,Wouters2017}. The systematic expansion in static bath orbitals of EwDMET is generalized for a systematic expansion of explicitly {\em dynamic} bath orbitals to converge the bath discretization error for fully dynamical properties. As with EwDMET, this again relies on the representation of the self-consistently obtained self-energy in an explicit auxiliary-space representation, such that the Weiss field can also be represented via diagonalization of a quadratic Hamiltonian form\cite{PhysRevB.89.035148}.
This allows for the use of Schmidt decomposition techniques on a single-particle wavefunction
to derive appropriate bath orbitals suitable for a DMFT-like framework
with explicit and fully dynamical effects
(in comparison to the limited dynamical resolution of EwDMET).
%
%
It may be argued that the construction of this auxiliary space representation of the self-energy is as challenging as the construction of the bath space in ED-DMFT. However in practice, this is a far less severe constraint than the original
bath orbital problem required to represent the hybridization, as the cost of inclusion of additional auxiliary self-energy states is just (static) mean-field cost, 
and so the self-energy does not require a compact representation in terms of these auxiliary states.

In this work, we will test the \wdmft{} method on different lattices
with the Hubbard~\hamiltonian{}
\begin{equation}
    H = h + H_U =
    -t \mkern-8mu \sum_{<ij>,\sigma} \mkern-6mu c_{i\sigma}^{\dagger} c_{j\sigma}
    + U \sum_{i} n_i^\up n_i^\down,
\end{equation}
where $t$ describes the nearest neighbor hopping and $U$ the on-site electron interaction.
Despite its simplicity, the Hubbard~\hamiltonian{} captures
the competing effects of delocalization due to the
kinetic energy term and localization due to the electron correlation.
Additionally, for special lattice geometries, like the infinite dimensional and one-dimensional
cases, exact solutions of some properties are known, which we will extensively compare against.
Throughout this paper all energies are understood in units of $t$.

\section{Theory} \label{sec:theory}
We first give a short description of DMFT, while we refer to
Refs.~\onlinecite{Georges1996} and \onlinecite{Kotliar2006}
for more in-depth reviews.
We then present in more detail the individual steps of the \wdmft{} method,
as well as the self-consistent algorithm and the calculation of total energy
and spectral function.

\subsection{DMFT}

%
The fundamental approximation of DMFT is the neglect of non-local, long-range contributions to the electronic self-energy.
Dividing a system of $N_\mathrm{sites}$ into $N_\mathrm{sites}/N_\imp$ fragments of size $N_\imp$,
which are related by some symmetry operation of the system,
all blocks of the self-energy coupling different fragments are assumed zero
by virtue of the DMFT approximation.
In lattices with translational symmetry, this can be expressed as
momentum-independence of the self-energy
as
$\Sigma(k,z) = \Sigma_\mathrm{imp}(z)$,
where $\Sigma(k,z)$ is the lattice self-energy of momentum $k$ at
frequency $z$ and $\Sigma_\mathrm{imp}$ is the impurity self-energy.
To calculate the latter, an impurity problem has to be set up
consisting of a single fragment coupled to its environment as given by the hybridization, which defines the single-particle character of this coupling.
In the Hamiltonian formulation of this problem, this coupling is approximated by a set of $N_{\mathrm{b}}$ bath orbitals, which describe
the effect of the frequency-dependent hybridization $\Delta(z)$ with the environment.
The energies $\epsilon_p$ and couplings $\Gamma_p$ (a vector) of each bath orbital describe the
cluster hybridization
\begin{equation}
\Delta_\mathrm{c}(z) =
\sum_{p}^{N_\mathrm{b}}
\frac{\Gamma_{p} \otimes \Gamma_{p}^*}
{z - \epsilon_p}
.
\end{equation}
Finding appropriate bath parameters $\epsilon_p$ and $\Gamma_p$ is one of the
major challenges of ED-DMFT.
In general, the process of finding an approximate $\Delta_\mathrm{c}(z)$ for
a given $\Delta(z)$
can be seen as a projection onto a functional subspace of all
possible hybridizations. \cite{Georges1996}
To perform this projection, one can introduce a distance function in the functional space and then numerically minimize this distance by varying all bath parameters.
%
If performed on the imaginary frequency axis, the minimization is generally well behaved, for example using the distance function
\begin{equation}
d =  \sum_n^{N_{\omega}} \frac{1}{\omega_n^M}
\tr \left| \Delta(\iw) - \Delta_\mathrm{c}(\iw) \right|^2
\label{eq:dmft-fit}
,
\end{equation}
where $N_{\omega}$ determines a frequency cutoff and $M$ a frequency weighting,
which is often chosen between 0 and 2 (in this work we always use all available
Matsubara points and $M = 1$).
Instead of taking the difference of the hybridizations in \eq{}\eqref{eq:dmft-fit},
it is also possible to use the corresponding Green's~functions instead.\cite{Liebsch2012}
The different ways of defining the distance function as well
as an often significant dependence on the starting guess for the fit parameters,
makes the fitting procedure an unsatisfying step of DMFT,
which often needs to be tuned to the physical problem at hand.
We note that alternative ways to obtain bath parameters exist, for example via
a real-frequency description of the hybridization required for numerical renormalization group solvers\cite{Pruschke08,Anders06}, 
truncation of a continued-fraction representation\cite{Si1994}, or further constraints on the numerical fitting to improve the numerics.\cite{HybFit}
%
%
Once the bath parameters are determined, the cluster \hamiltonian{}
\begin{align}
H_\mathrm{c} =
    \sum_{ab \in \imp} h_{ab} c_a^\dagger c_b
    + U \sum_{a \in \imp} n_a^\up n_a^\down
    + \sum_{p \in \mathrm{bath}}  \epsilon_p c_p^\dagger c_p \nonumber \\
    + \sum_{a \in \imp} \sum_{p \in \mathrm{bath}}
    \left( \Gamma_{ap} c_a^\dagger c_p + \mathrm{h.c.} \right)
\end{align}
can be solved for the ground-state $\ket{0}$ and its energy $E_0$ via exact diagonalization.
Note that we suppress spin indices and summations for simplicity, except for the interaction
term where they are denoted explicitly.
The impurity Green's function
\begin{align}
G_{\imp,ab}(\iw) =
& \braket{0 | c_a \left[ \iw - (H_\mathrm{c} - E_0) \right]^{-1} c_b^\dagger | 0} \\
+ & \braket{0 | c_b^\dagger \left[ \iw + (H_\mathrm{c} - E_0) \right]^{-1} c_a | 0}
\end{align}
can be calculated via a number of different methods, traditionally the dynamical Lanczos approach\cite{Prelov2013}. However in this work, we use a more accurate approach which computes the Green's function exactly, one frequency at a time, via the correction vector method\cite{Soos1989,Kuhner1999}.
The impurity self-energy is then defined by the Dyson equation
\begin{equation}
\Sigma_\mathrm{imp}(\iw) =
\iw - h_\mathrm{imp} - \Delta_\mathrm{c}(\iw)
- G_\mathrm{imp}^{-1}(\iw)
\label{eq:dmft-s-imp}
,
\end{equation}
where $h_\mathrm{imp}$ is the projection of the one-electron \hamiltonian{} into the impurity space.
The impurity self-energy can be unfolded into the lattice self-energy
using the symmetry operations of the underlying system.
Labelling the equivalent fragments with $R$ and $R'$, this can be written as
\begin{equation}
\Sigma_{RR'}(\iw) = \delta_{R R'} \Sigma_\mathrm{imp}(\iw)
.
\label{eq:dmft-replicatese}
\end{equation}
The lattice self-energy defines the lattice Green's~function
$G(\iw) = \left[ \iw - h - \Sigma(\iw) \right]^{-1}$,
which in turn defines a new hybridization function
\begin{equation}
\Delta(\iw) =
\iw - h_\mathrm{imp}
- \left[ G_{00}(\iw) \right]^{-1} - \Sigma_{00}(\iw)
\label{eq:dmft-hybrid}
,
\end{equation}
where $G_{00}$ and $\Sigma_{00}$ refer to the impurity part of the lattice
Green's function and self-energy, respectively.
From the new hybridization, an updated set of
bath parameters $\epsilon$ and $\Gamma$ can be fitted via the minimization of
Eq.~\eqref{eq:dmft-fit}.
This procedure is then iterated until self-consistency is achieved.
An advantage of the ED solver in DMFT is that the
ill-conditioned analytic continuation from Matsubara Green's function
to the real axis is circumvented.
Instead, once the bath parameters are converged on the Matsubara axis,
the impurity Green's function can be calculated at real frequencies.
As a consequence of the bath discretization error,
the spectral function obtained this way will in generally not be smooth,
as it will necessarily be represented as a sum of a finite number of
individual poles of the impurity model.
Alternatively, one can calculate the self-energy on the real-frequency
axis, and then subsequently compute the spectral function
in the lattice space, according to
\begin{equation}
A(\omega) = -\frac{1}{\pi} \Im G_{00}^\mathrm{R}(\omega)[\Sigma(\omega)]
\label{eq:dmft-a-lat}
,
\end{equation}
where $G^\mathrm{R}$ is the retarded Green's function of the lattice.
This will remove the explicit bath discretization error in the one-particle \hamiltonian{} and therefore yield smoother spectra, however it will still potentially suffer from the implicit error due to the bath discretization effects in the self-energy.
With the exception of \fig{}\ref{fig:g-cluster}, the DMFT spectral functions shown in
this paper are calculated according to Eq.~\eqref{eq:dmft-a-lat}.

\subsection{\wdmft{}}

Since a compact and robust bath parameterization via numerical minimization of
\eq{}\eqref{eq:dmft-fit} is a major difficulty in DMFT calculations,
we avoid it by constructing bath orbitals algebraically from an expression
that is derived from the Schmidt decomposition of a
mean-field response wavefunction, given by
\begin{equation}
\ket{\phi^{(1)}_a (z)} = \frac{1}{z-h} c^{(\dagger)}_a \ket{\Phi_0}
\label{eq:response}
,
\end{equation}
where $h$ is a quadratic \hamiltonian{} which can implicitly include the effects of a self-consistent self-energy, 
thereby generalizing the work of Ref.~\onlinecite{Booth2015}.
The bath orbitals will thus inherit the frequency-dependence of the response wavefunction, ultimately resulting in a different cluster problem for each frequency.
In contrast to standard DMFT, \wdmft{} will guarantee that the uncorrelated Green's~function is exactly reproduced in the impurity space (i.e. in the absence of
explicit two-electron terms).
As a consequence of the Dyson~equation, the hybridization is therefore also exactly matched.
The Schmidt decomposition technique comes with the caveat that only single Slater-determinant wavefunctions can be decomposed into one-electron bath orbitals.
Therefore, as an additional difference to standard DMFT, it is thus necessary to
augment the system with auxiliary degrees of freedoms,
which describe the effect of the dynamical self-energy
on the lattice, whilst ensuring that $h$ in \eq{}\ref{eq:response}
remains a quadratic Hamiltonian.
In the following subsections we describe the derivation of the bath orbitals as well as the other
steps of the \wdmft{} method in more detail.

%

\subsubsection{Bath construction}\label{sec:bath}
%
The wavefunctions which we aim to Schmidt decompose in order to find the appropriate
bath states can be written in their
most general form as $f(H) c_a^{(\dagger)} \ket{0}$ where $f$
is a function of the \hamiltonian{} $H$.
In this work we restrict ourselves to a \hamiltonian{} which is quadratic
in the electron-field, so that its true ground-state
$\ket{0}$ can be expressed as a single Slater-determinant $\ket{\Phi_0}$.
The bath wavefunctions then become simple tensor products of
single bath orbitals and a determinant of the remaining orbitals in $\ket{\Phi_0}$\cite{Wouters2016}.
Many-body effects via an implicit self-energy can still be retained via an augmented 
one-electron \hamiltonian{} $h_\weiss$, which we will define in Sec.~\ref{sec:aux}. In the first iteration (i.e. in the absence of an effective self-energy), $h_\weiss=h$.

\begin{figure}
\includegraphics[width=\linewidth]{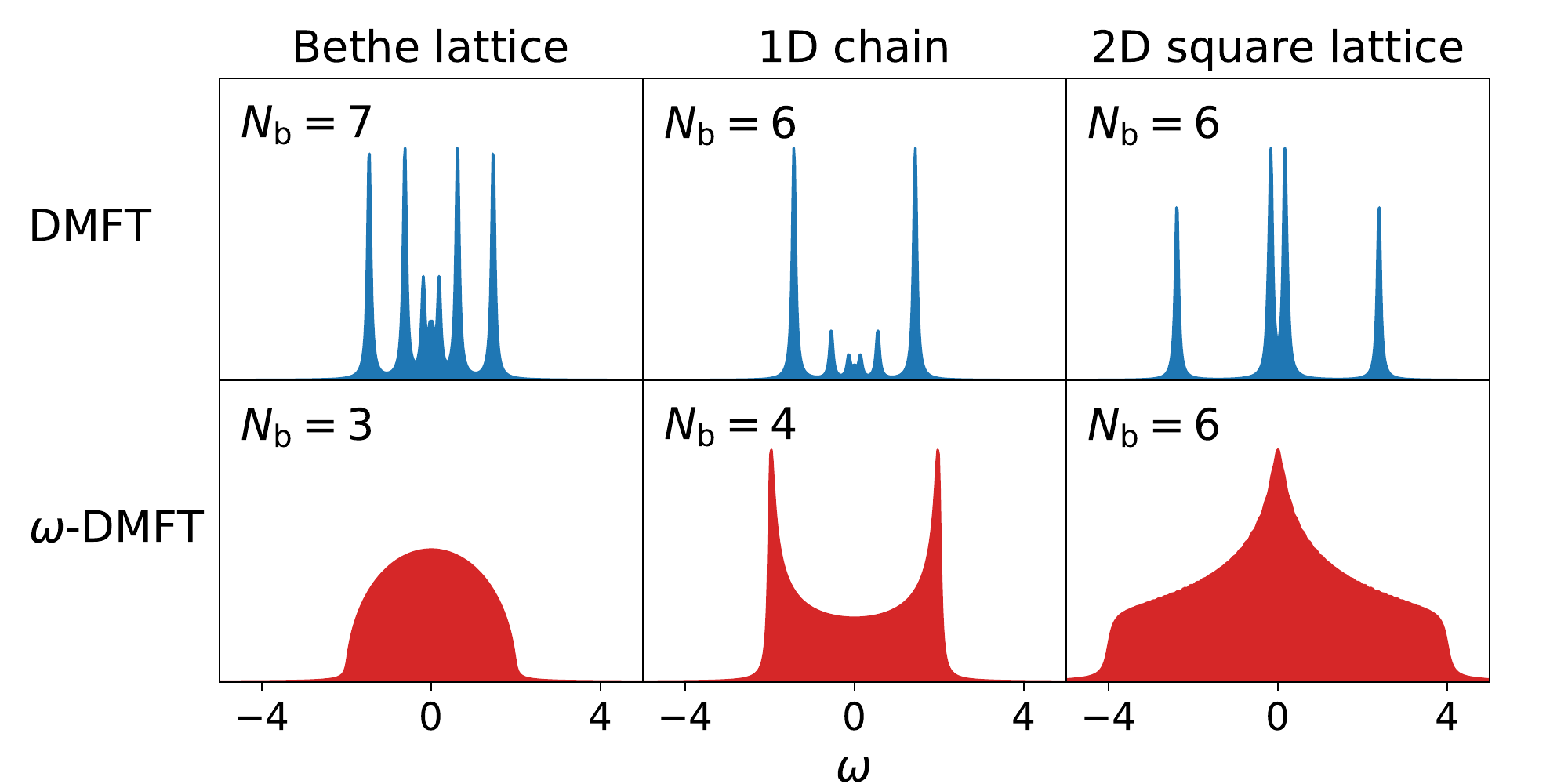}
\caption{
    \label{fig:g-cluster}
    Impurity Green's~function of the auxiliary model for DMFT
    and the \wdmft{} method for different lattice geometries
    in the absence of explicit two-body impurity interactions.
    The introduction of frequency-dependent bath orbitals in \wdmft{} means
    that the impurity Green's function in the auxiliary system exactly
    matches the desired lattice spectrum, resulting in
    an exact hybridization and smooth spectral functions,
    despite a small finite bath space of $N_\mathrm{b}$ orbitals.
}
\end{figure}

\begin{figure}
\includegraphics[width=\linewidth]{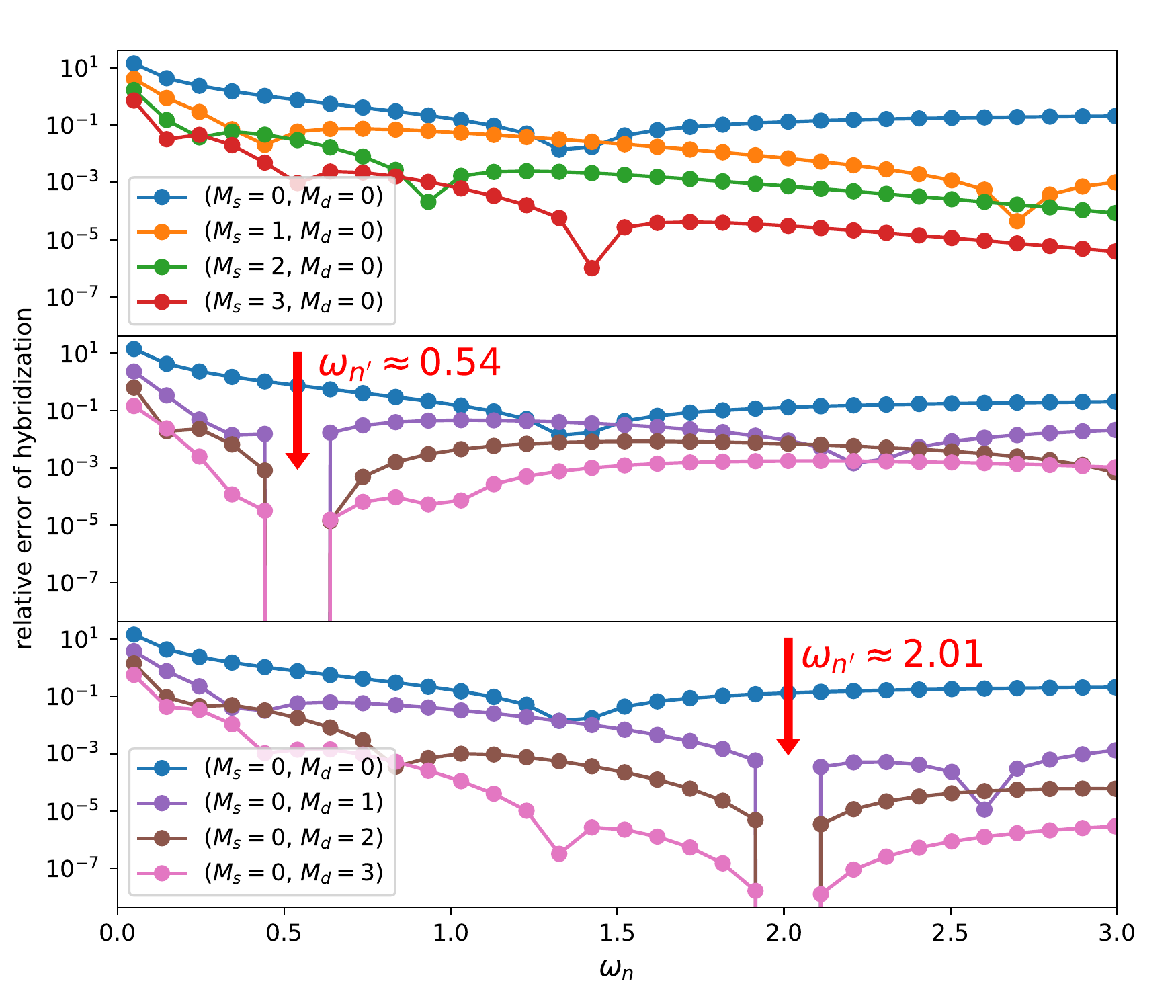}
\caption{\label{fig:bethe-hybrid}
Relative error of hybridization function,
defined as $\left( \Delta_\mathrm{c}(\iw) - \Delta(\iw) \right) / \Delta(\iw)$,
for the Hubbard model on the Bethe lattice at $U=4$.
The maximum order of static bath orbitals is denoted $M_\mathrm{s}$
and the maximum order of dynamic bath orbitals $M_\mathrm{d}$.
Top panel: Convergence with respect to $M_\mathrm{s}$ without any dynamic bath orbitals.
Middle panel: The dynamic bath orbitals are evaluated at the 6th Matsubara frequency (red arrow),
and the hybridization is matched exactly at this frequency.
Bottom panel: The same as the middle panel,
but with the dynamic bath orbitals evaluated at the 21st Matsubara frequency.
It is clear that matching higher-order derivatives of the hybridization by
increasing $M_\mathrm{d}$ at a given frequency point also improves the
hybridization error away from the frequency of interest.
}
\end{figure}

In EwDMET, the decomposition is done on the wavefunction
$\left(h_\weiss\right)^m c_a^{(\dagger)} \ket{\Phi_0}$, which leads to the particle and hole bath orbitals
\begin{alignat}{2}
\ket{b_{a,m}^{>}} &=
\sum_{x \notin \textrm{imp}}
\sum_{i : \epsilon_i > \mu}
C_{ai} \epsilon_i^m C_{xi} \ket{x} &&\quad m \geq 0
\label{eq:stat-bath-p}
\\
\ket{b_{a,m}^{<}} &=
\sum_{x \notin \textrm{imp}}
\sum_{i : \epsilon_i < \mu}
C_{ai} \epsilon_i^m C_{xi} \ket{x} &&\quad m > 0,
\label{eq:stat-bath-h}
\end{alignat}
where $C$ and $\epsilon$ are the eigenvectors and eigenvalues of $h_\weiss$.\cite{Fertitta2018}
We will call these orbitals {\em static} bath orbitals to distinguish them from the frequency-dependent
dynamic bath orbitals, which will be introduced below.
The standard DMET bath orbitals are the special case $m = 0$ of these orbitals.
Note that in this $m = 0$ case the particle and hole orbitals are linearly dependent
and the hole orbitals can thus be restricted to orders greater than zero.
The particle static bath orbitals ensure that the moments of the particle (unoccupied) spectral function, 
$T_{ab,n}^> = \sum_{i:\epsilon_i > \mu} C_{ai} \epsilon_{i}^n C_{bi}$, are matched between $h_\weiss$
and its projection into the impurity+bath space, and similarly for the hole (occupied) bath orbitals.
The maximum moment~$n$ that will be exactly matched by including all bath orbitals
up to order $M_\mathrm{s}$ is given by $n = 2M_\mathrm{s} + 1$,
where $M_\mathrm{s}$ denotes the maximum exponent $m$ in
Eqs.~\ref{eq:stat-bath-p} and \ref{eq:stat-bath-h}.\cite{Fertitta2018}
Since the moments relate directly to the high frequency expansion of the Green's function~
$\left(z - h_\weiss\right)^{-1}$, adding static bath orbitals corresponding to higher moments can be seen as
a systematic expansion of the Green's function in the high-frequency limit,
though the fact that particle and hole moments are separately matched
means that this is not purely a high-energy expansion.
In order to match the Green's function between lattice and cluster space,
we apply the Schmidt decomposition directly to frequency-dependent wavefunction
$\left(z - h_\weiss\right)^{-1} c_a^{(\dagger)} \ket{\Phi_0}$,
which was proposed in Ref.~\onlinecite{Booth2015}.
The advantage of using frequency-dependent bath orbitals constructed this way is exemplified in
\fig{}\ref{fig:g-cluster}, where the impurity density of states (DoS) of the auxiliary impurity+bath system is compared
between standard DMFT and \wdmft{} at $U = 0$.
While the bath discretization is very apparent in the DMFT case,
the dynamic bath orbitals of \wdmft{} guarantee that the
uncorrelated Green's function is exact at every frequency, even in the presence of
an effective self-energy in the lattice.
To additionally match frequency-derivatives of the Green's~function at $z$,
we can generalize the construction above and decompose
$\left(z - h_\weiss\right)^{-m} c_a^{(\dagger)} \ket{\Phi_0}$ instead, with $m = 1$
representing the previous case.
The Schmidt decomposition of this wavefunction results in the dynamic bath orbitals
\begin{alignat}{2}
\ket{f_{a,m}^{\mathrm{>}}(z)} &=
\sum_{x \notin \textrm{imp}}
\sum_{i: \epsilon_i > \mu}
\frac{C_{ai} C_{xi}}
{(z -\epsilon_i)^m}
\ket{x} &&\quad m > 0
\label{eq:dyn-bath-p}
\\
\ket{f_{a,m}^{\mathrm{<}}(z)} &=
\sum_{x \notin \textrm{imp}}
\sum_{i: \epsilon_i < \mu}
\frac{C_{ai} C_{xi}}
{(z -\epsilon_i)^m}
\ket{x} &&\quad m > 0
\label{eq:dyn-bath-h}
.
\end{alignat}
One might expect that every order of dynamic bath orbitals matches two
orders of derivatives of the Green's~function, similar how every static bath order
ensures the matching of two additional moments.
However, this is only the case if also the complex conjugated counterparts
to Eqs.~(\ref{eq:dyn-bath-p},\ref{eq:dyn-bath-h}) are added to the bath space.
This would then result in the rule $n = 2 M_\mathrm{d} - 1$, where $n$ is the maximum derivative
that is matched (with $n = 0$ referring to the value of the hybridization at frequency~$z$)
and $M_\mathrm{d}$ is the maximum order $m$ for which the dynamic bath orbitals and their
complex conjugates are included.
Equivalently one can add the real and imaginary part of the dynamic bath orbitals
individually, spanning the same bath space
but avoiding complex arithmetic altogether.
While we believe that this is the most efficient way to match higher
derivatives of the hybridization,
in this work we only use the complex dynamic bath orbitals directly
and the maximum matched derivative is simply $n = M_\mathrm{d} - 1$.
The full frequency-dependent bath space $\mathcal{B}(z)$ is constructed as the tensor product
of the individual static and dynamic bath orbitals as
\begin{equation}
\mathcal{B}(z) =
\bigotimes_{a \in \imp}
\left[
\ket{b_{a,0}^<}
\hspace{-0.5em}
\bigotimes_{\zeta \in \{>,<\}}
\hspace{-0.3em}
\left(
\bigotimes_{m = 1}^{m \leq M_\mathrm{s}}
\ket{b_{a,m}^\zeta}
\bigotimes_{m = 1}^{m \leq M_\mathrm{d}}
\ket{f_{a,m}^\zeta(z)}
\right)
\right]
\label{eq:wdmft-bath}
,
\end{equation}
where $M_\mathrm{s}$ ($M_\mathrm{d}$) denotes the maximum order of static (dynamic) bath orbitals.
We note that the orbitals in Eqs.~(\ref{eq:stat-bath-p}-\ref{eq:dyn-bath-h}) are not a convenient
basis for $\mathcal{B}$, as they are not generally orthonormal.
We thus perform a QR~decomposition on the matrix representing $\mathcal{B}$
in order to obtain an orthonormalized set of bath orbitals.
While the inclusion of the first-order ($m = 1$) dynamic bath orbitals guarantees
an exact matching of
the hybridization, we want to emphasize that this does not completely remove the
bath discretization error.
To understand this, it is useful to add a parameterization in terms of
an additional frequency $z'$ to Eq.~\eqref{eq:dmft-hybrid}:
\begin{equation}
\Delta_\mathrm{c}(z;z') = \sum_p^{N_\mathrm{b}} \frac{\Gamma_p(z') \otimes \Gamma_p^*(z')}{z - \epsilon_p(z')}
.
\end{equation}
What we mean with \textit{exact matching of the hybridization} is that the
bath parameterization at $z'$ guarantees that the hybridization is exact at
$z = z'$, but not at other frequencies $z \neq z'$.
Similarly, the higher order dynamic bath orbitals guarantee that the
first, second, etc derivatives with respect to $z$ are exact at $z'$.
This is illustrated in Fig.~\ref{fig:bethe-hybrid},
where the relative error of the cluster hybridization
$\Delta_\mathrm{c}(\iw;\iu \omega_{n'})$ is plotted as a function of $\iw$
for the two sets of bath orbitals parameterized for the fixed frequency points $n' = 5$ (middle panel) and $n' = 20$ (bottom panel).
This is compared to the performance of the expansion in higher order {\em static}
bath orbitals of EwDMET (top panel).
While the static bath orbitals quickly reduce the hybridrization error in the
high frequency regime, the error at low frequencies reduces only slowly with $M_\mathrm{s}$.
On the other hand, the dynamic bath orbitals lead to an exact matching of the hybridization at the point they are meant to represent, while the higher order dynamic bath orbitals also construct an increasingly good match of the hybridization around these points.

Why is it important to match the hybridization at frequencies other than the frequency which is explicitly being solved for with the given bath orbitals? This is because the introduction of the explicit two-particle impurity terms in the impurity model couples the hybridization at different frequency points. It is a reasonable assumption that the main relevant energy scale will be centered around the frequency point of interest, and therefore, matching the higher-order derivatives of the hybridization at the frequency point of interest seems a reasonable strategy, and one which is borne out from numerical results.
Therefore, in this work we will not consider higher order static bath orbitals and always
choose $M_\mathrm{s} = 0$, and choose to systematically expand the bath space in
the presence of explicit interactions by increasing the maximum dynamic bath orbital order.
This simplifies \eq{}\eqref{eq:wdmft-bath} to
\begin{equation}
\mathcal{B}(z) =
\bigotimes_{a \in \imp}
\left[
\ket{b_{a,0}^>}
\hspace{-0.5em}
\bigotimes_{\zeta \in \{>,<\}}
\hspace{-0.3em}
\bigotimes_{m = 1}^{m \leq M_\mathrm{d}}
\ket{f_{a,m}^\zeta(z)}
\right]
\label{eq:wdmft-bath-ms0}
\end{equation}
and reduces the number of bath orbitals to $N_\mathrm{b} = N_\mathrm{imp} (1 + 2M_\mathrm{d})$.
For ED solvers, this may still result in a prohibitively large number of bath orbitals for a
larger number of impurity sites, even for the minimal case of $M_\mathrm{d} = 1$.
We thus also implement the following factorization of the dynamic bath orbital space:
to calculate a column $b$ of the particle Green's matrix $G_{ab}^>$, only
the dynamic bath orbitals constructed from $b$, i.e. $\ket{f_{b,m}^>(z)}$ and
$\ket{f_{b,m}^<(z)}$ are included in the bath space.
This can similarly hold for the rows $a$ of the hole Green's function.
This makes the bath space of \eq{}\eqref{eq:wdmft-bath-ms0} not only frequency,
but also impurity-orbital dependent:
\begin{equation}
\mathcal{B}_b(z) =
\bigotimes_{a \in \imp}
\ket{b_{a,0}^>}
\bigotimes_{\zeta \in \{>,<\}}
\bigotimes_{m = 1}^{m \leq M_\mathrm{d}}
\ket{f_{b,m}^\zeta(z)}
\label{eq:wdmft-bath-orbfac}
.
\end{equation}
We call this approach `orbital factorization', as it only matches the hybridization
(and optionally its derivatives) for a single impurity orbital at a time.
The number of bath orbitals (for any given cluster) then becomes
$N_\mathrm{b} = N_\mathrm{imp} + 2M_\mathrm{d}$, which makes the calculation of e.g. a
complete $d$-shell feasible within ED for a $M_\mathrm{d} = 1$ bath, resulting in
7 bath orbitals and a cluster size of 12.
Note that an impurity-orbital dependent bath space could potentially lead to a breaking of the
symmetry $G_{\imp,ab}(\iw) = G_{\imp,ba}(\iw)$ and require an additional symmetrization step.
In the systems tested in this work, however, this symmetry was not broken.
%
%
We note that this factorization is exact in
the limit of $U = 0$, retaining the beneficial feature of an exact representation of
the uncorrelated Green's~function in the cluster space.
Another issue in standard DMFT is the electron filling of the cluster problem.
In general, separate calculations with different fillings are necessary to determine
the best choice, usually by selecting the filling which results in the lowest ground-state
energy of the impurity model.
This problem is not present in \wdmft{}: the zero-order static bath orbitals have a filling
of $2 - \chi_\alpha$ where $\chi_\alpha$ is the filling of the impurity orbital which they
are derived from according to \eq{}\eqref{eq:stat-bath-p}.
Consequently, the impurity + zero-order static bath space can always be filled with
a known, integer number of electrons.
This property still holds when higher order static or dynamic bath orbitals are added,
since these are always completely empty (particle orbitals)
or completely filled (hole orbitals).
As a result, the electron number of the cluster problem is known by construction and exactly
matches the electron number of the projection of the lattice density matrix into
the cluster space.

To conclude this section, it is worth making a connection between the ideas in this approach and the method of distributional exact diagonalization (DED).\cite{Granath2012}
In DED, finite bath spaces for the impurity problem are sampled stochastically according to
a probability distribution derived from the continuous hybridization, and the self-energy is obtained as a sample average.
This corresponds to a neglect of a subset of the bath-bath cross correlation terms
in the diagrammatic self-energy expansion.
A similar effect occurs in the bath discretization choice of \wdmft{}. Here, there are a set of cluster problems,
each constructed to best represent the hybridization effects with the environment
at a given frequency. However, bath discretization errors can still emerge once interacting terms are included in the impurity Hamiltonian, as this can couple the hybridization at different frequency points.
The final impurity self-energy is then a summation of the individual contributions,
effectively neglecting some of the cross-frequency correlations in the
self-energy expansion. However, this inter-frequency coupling is systematically converged via the addition of bath orbitals derived from the Schmidt decomposition of higher-order dynamical functions.

\subsubsection{Lattice self-energy}\label{sec:aux}
As mentioned above, many-body effects can be included in the lattice to describe the correlated
changes to the one-particle environment of the impurity. This has an effect on the construction
and nature of the bath states, and is described in DMFT via the self-consistent local self-energy.
In \wdmft{}, we replace $h$, the one-particle part of the lattice
Hamiltonian, with an augmented one-particle \hamiltonian{} $h_\mathrm{Weiss}$, which is designed to
represent the Weiss field of the impurity space, implicitly including the self-energy effects in the environment.
In DMET this is attempted via a simple one-electron potential which cannot describe dynamical
correlation effects, while in EwDMET and in this work, we
additionally use a set of auxiliary states which are coupled to the system.
The benefit of the latter method is that the auxiliary states are able to
represent the poles of any arbitrary frequency-dependent self-energy.
With the energies $\alpha$ and couplings $\beta$ of the auxiliary states,
the augmented Weiss-\hamiltonian{} is
\begin{multline}
h_\mathrm{Weiss} =
    \sum_{ij \in \mathrm{lat}} h_{ij} c_i^\dagger c_j
    + \sum_{F > 0} \left[
    \sum_{ij \in \mathcal{F}(F)} \Sigma^{\infty}_{ij} c_i^\dagger c_j
    + \sum_{r \in \mathcal{A}(F)} \alpha_r c_r^\dagger c_r
        \right. \\ \left.
        + \sum_{i \in \mathcal{F}(F)} \sum_{r \in \mathcal{A}(F)}
        \beta_{ir} \left( c_i^\dagger c_r + \mathrm{h.c.} \right)
        \right]
    \label{eq:wdmft-h-weiss}
    ,
\end{multline}
where $F$ is an index that enumerates the symmetry equivalent fragments
of the lattice
with $F = 0$ being the impurity space and $\mathcal{F}(F)$ is the set of sites in $F$. This achieves the same
replication of the self-energy in DMFT given by Eq.~\eqref{eq:dmft-replicatese}.
$\mathcal{A}(F)$ denotes a subspace of the full auxiliary space, which contains
auxiliaries only coupling to $F$.
By virtue of the DMFT approximation (i.e. an impurity-local effective self-energy), no auxiliaries 
that couple between fragments are needed in Eq.~\eqref{eq:wdmft-h-weiss}.
Additionally, the impurity space is excluded from the summation in
Eq.~\eqref{eq:wdmft-h-weiss}, which is in contrast to DMET,
where the correlation potential
(taking the role of the static self-energy $\Sigma^\infty$)
is also included in the impurity space in the definition of the Hamiltonian to perform the Schmidt decomposition.
%
%
This can be compared to DMFT, where the hybridization is also defined with respect
to an uncorrelated impurity, as the self-energy is removed from the impurity in the final term in
Eq.~\eqref{eq:dmft-hybrid}.
Only the removal of the impurity space auxiliaries (and static self-energy)
leads to a systematic convergence to the correct hybridization
with higher orders of static or dynamic bath orbitals. We note here that previous (static) EwDMET work did not
remove this local effective self-energy in the bath construction, and the effect of this will be
investigated in future work.
%
%
%

%
To determine an appropriate set of auxiliary parameters $\alpha$ and $\beta$
from the impurity self-energy, we minimize the distance functional
\begin{equation}
d = \sum_n \frac{1}{\omega_n}
\tr \left| \Sigma_\mathrm{aux}(\iw;\alpha\beta)
- \left( \Sigma_\mathrm{imp}(\iw) - \Sigma_\mathrm{imp}^\infty \right) \right |^2
\label{eq:semb-fit}
,
\end{equation}
by variation of $\alpha$ and $\beta$ in the auxiliary self-energy
\begin{equation}
\Sigma_\mathrm{aux}(\iw;\alpha\beta) =
\sum_{r}^{N_\mathrm{aux}} \frac{\beta_r \otimes \beta_r}
{\iw - \alpha_r}
\label{eq:semb-s-aux}
.
\end{equation}
Note that this step is equivalent to the bath fit via minimization of
\eq{}\eqref{eq:dmft-fit} in DMFT.
The only difference between bath states and auxiliary states is the
physics they represent:
the bath represents the delocalization and one-electron hybridization
with the environment, whereas the auxiliaries above represent
many-body correlation effects due to a self-energy within the impurity.
While it seems that the bothersome non-linear fit of the bath orbitals
from DMFT is just replaced with another fit of this kind in \wdmft{},
it has to be stressed that the auxiliaries are added to the
mean-field \hamiltonian{} and not to the cluster problem.
The computational complexity thus only increases as
$\mathcal{O}(N_\mathrm{aux}^3)$ instead of exponentially and
converging the calculations with respect to the number of
auxiliaries was not a problem for the systems in this paper, as the emphasis on a compact representation is not necessary.
In practice we find that 20~auxiliary states per impurity orbital,
initially distributed particle-hole symmetrically on a logarithmic energy grid
between $-10 t$ and $10 t$ are sufficient.
The initial elements of the coupling vectors were set to $0.01 t$ or $-0.01 t$,
depending on to which sublattice of the bipartite lattice they couple to, respectively
(for a calculation with a single impurity site, all elements were initially positive).
It should be noted that while more auxiliaries can be added, the fit can become
overcomplete, since it is performed on the Matsubara axis
and many solutions can exist that minimize the difference function to a
similar level of accuracy.\cite{Lu2017}
%

\subsubsection{Solving the impurity problem}
To simplify the notation in the following, we introduce the index $\kappa(z)$,
which enumerates the frequency-dependent impurity problems to be solved at each frequency
(this index must not be confused with the fragment index $F$
in \eq{}\eqref{eq:wdmft-h-weiss}, which labels the symmetry-equivalent impurity clusters in the lattice).
If the impurity-orbital factorization is used, $\kappa$
additionally depends on the impurity orbital.
%
%
Denoting the projector into the space of $\kappa$ as $P_\kappa$,
we can then write down the embedded cluster \hamiltonian{}, equivalent to the auxiliary Anderson impurity model of DMFT, as
\begin{equation}
H_\kappa = P_\kappa h_\mathrm{Weiss} P_\kappa
+ U \sum_{a \in \mathrm{imp}} n_a^\up n_a^\down
- \frac{U}{2} \sum_{a \in \imp} \gamma_{aa} c_a^\dagger c_a
\label{eq:semb-h-cl}
,
\end{equation}
%
where the last term subtracts the Hartree--Fock potential, which is already contained in
$h_\weiss$ (for simplicity we suppressed both the Hartree--Fock potential
and the chemical potential term in \eq{}\eqref{eq:wdmft-h-weiss}, since they cancel exactly).
We note that in the absence of particle-hole symmetry it becomes necessary to introduce
an energy shift between impurity sites and bath sites in \eq{}\eqref{eq:semb-h-cl},
which needs to be optimized to ensure the correct distribution of
electrons between impurity and bath.
For every cluster $\kappa$, we use exact diagonalization to solve
\begin{equation}
H_\kappa \ket{0_\kappa} = E_{0;\kappa} \ket{0_\kappa},
\label{eq:semb-gs}
\end{equation}
%
%
for the ground-state and LGMRES\cite{Baker2005} 
to solve the particle and hole response equations
\begin{align}
\left[\iw - (H_\kappa^> - E_{0;\kappa}) \right]
\ket{\phi_{a;\kappa}^>(\iw)} &= c_a^\dagger \ket{0_\kappa}
\label{eq:semb-rs-p} \\
\left[\iw + (H_\kappa^< - E_{0;\kappa}) \right]
\ket{\phi_{a;\kappa}^<(\iw)} &= c_a \ket{0_\kappa}
\label{eq:semb-rs-h}
\end{align}
for the correction vectors $\ket{\phi_{a;\kappa}^>(\iw)}$ and $\ket{\phi_{a;\kappa}^<(\iw)}$.
Note that we cannot solve solely for the imaginary part of the correction vector
as is usually done using the conjugate gradient method, because the \hamiltonian{} is complex.
\cite{Soos1989,Kuhner1999}
This could be circumvented in the case of including independent real and imaginary parts of the dynamic bath orbitals
to ensure a real \hamiltonian{}.
With the ground-state and both response states calculated, the impurity
Green's~function is simply
\begin{equation}
G_{\mathrm{imp},ab}(\iw) =
  \braket{0_\kappa | c_a | \phi_{b;\kappa}^>(\iw)}
+ \braket{\phi_{a;\kappa}^<(\iw) | c_b | 0_\kappa}
\label{eq:semb-g-imp},
\end{equation}
and the impurity self-energy is given by the Dyson equation
\begin{equation}
\Sigma_\mathrm{imp}(\iw) =
\pimp \left[ \iw - h_\mathrm{Weiss} \right]^{-1} \pimp - G_\mathrm{imp}^{-1}(\iw)
\label{eq:semb-dyson},
\end{equation}
where $\pimp$ is a projector into the impurity space.
Note that in Eq.~\eqref{eq:semb-dyson}, we could have used the cluster
projection instead of the full~$h_\mathrm{Weiss}$ (i.e. $P_\kappa h_\mathrm{Weiss} P_\kappa$).
This is because the uncorrelated Green's~function will always truthfully
be represented in the cluster space due to the first-order
dynamic bath orbitals.
In DMFT on the other hand, these two choices differ, the first corresponding to
using the true hybridization $\Delta(\iw)$ for the bath Green's function in
Eq.~\eqref{eq:dmft-s-imp} and the latter to using the
fitted cluster hybridization~$\Delta_\mathrm{c}(\iw)$
(In practice, in DMFT $\Delta_\mathrm{c}(\iw)$ is used for a better cancellation of bath discretization error between
the Weiss and impurity Green's function).

\subsection{Self-consistent Algorithm}
With the main steps of a \wdmft{} described above,
we now summarize the self-consistent algorithm.
The method starts from an initial guess for the impurity self-energy,
which can be taken from a different method,
a different model (e.g. a converged self-energy at a slightly different $U$ value), or if no such self-energy is available, can be set to zero.
The steps of the algorithm are then as follows:

\begin{enumerate}
\item Fit the auxiliary parameters $\alpha$, $\beta$ to the
impurity~self-energy via minimization of Eq.~\eqref{eq:semb-fit}.
\item Construct and diagonalize $h_\mathrm{Weiss}$~\eqref{eq:wdmft-h-weiss} to
obtain the eigenvalues $\epsilon$ and eigenvectors $C$.
\item Construct static and dynamic bath orbitals according to
Eqs.~(\ref{eq:stat-bath-p}-\ref{eq:dyn-bath-h}).
Orthonormalize the resulting bath matrix.
\item For each cluster $\kappa$,
solve the ground-state equation~\eqref{eq:semb-gs} with ED and
the response equations~(\ref{eq:semb-rs-p},\ref{eq:semb-rs-h}).
From the ground and response states, construct the
impurity Green's~function~\eqref{eq:semb-g-imp}
and the new impurity~self-energy via the
Dyson~equation~\eqref{eq:semb-dyson}.
\item To improve the stability of the algorithm,
we mix the old and new self-energy
according to
$\Sigma_\mathrm{imp} =
(1-\alpha_\mathrm{mix}) \Sigma_\mathrm{imp}^\mathrm{new}
+ \alpha_\mathrm{mix} \Sigma_\mathrm{imp}^\mathrm{old}$
with $\alpha_\mathrm{mix} = 0.25$.
\end{enumerate}
These steps are iterated, until the calculation converges.
As convergence criteria, we check that
\begin{multline}
\left\| \Sigma_{\mathrm{imp},ab}(\iw)
- \Sigma_{\mathrm{imp},ab}^\mathrm{old}(\iw) \right\|_\mathrm{F}
\\
< \left(1 - \alpha_\mathrm{mix} \right) \left( 10^{-4} + 10^{-3}
\left\| \Sigma_{\mathrm{imp},ab}(\iw) \right\|_\mathrm{F} \right)
\forall n,a,b
,
\end{multline}
where $\left\|\dots\right\|_\mathrm{F}$ denotes the Frobenius norm.

\subsection{Spectral function}
%
%
In ED-DMFT it is possible to get the real-frequency spectrum
after the bath parameters are converged on the Matsubara axis.
In principle, the same approach could be used in \wdmft{}, however in
contrast to DMFT, the dynamic bath orbitals in
Eqs.~(\ref{eq:dyn-bath-p},\ref{eq:dyn-bath-h})
would change with the transformation $z = \iw \to \omega + \iu \eta$.
Unfortunately, performing \wdmft{} on the real axis results in little control over
the sum rules in a frequency-factorized formulation
(even though we find that in practice the spectra obtained in
this manner are generally quite close to being normalized).
%
%
We thus choose instead to calculate the spectrum from the auxiliary model
of the self-energy, which must yield a causal and normalized
spectral function by construction, as the self-energy is represented in the form
of Eq.~\ref{eq:semb-s-aux}.
The mean-field \hamiltonian{} used for this is
\begin{multline}
h_\mathrm{aux} =
    \sum_{ij \in \mathrm{lat}} h_{ij} c_i^\dagger c_j
    + \sum_{F = 0} \left[
    \sum_{ij \in \mathcal{F}(F)} \Sigma^{\infty}_{ij} c_i^\dagger c_j
    + \sum_{r \in \mathcal{A}(F)} \alpha_r c_r^\dagger c_r
        \right. \\ \left.
        + \sum_{i \in \mathcal{F}(F)} \sum_{r \in \mathcal{A}(F)}
        \beta_{ir} \left( c_i^\dagger c_r + \mathrm{h.c.} \right)
        \right],
    \label{eq:semb-h-spec}
\end{multline}
which differs from the Weiss \hamiltonian{}~\eqref{eq:wdmft-h-weiss} only by inclusion of the
auxiliaries coupling to the impurity, i.e. the summation over $F$ includes $F = 0$, as the implicit self-energy
is now included in the impurity space.
The impurity part of the spectral function is then simply given by
\begin{equation}
A(\omega) = - \frac{1}{\pi} \Im \pimp \frac{1}{\omega - h_\mathrm{aux} + \iu \eta} \pimp.
\end{equation}
Note that this is equivalent to calculating the spectral function
in the lattice space in DMFT according to \eq{}\eqref{eq:dmft-a-lat}.

\subsubsection{Total energy}
The total energy per site of the DMFT and \wdmft{} method
can be calculated according to the Galitskii-Migdal equation
\cite{Galitskii1958}
\begin{align}
E = \frac{1}{N_\mathrm{imp}} \left[
\sum_{a \in \mathrm{imp}} \sum_{i \in \mathrm{lat}}
\left( h_{ai} + \delta_{ia} \frac{\gamma_{aa} U}{4} \right) \gamma_{ia}
\right. \nonumber \\ \left.
+ \sum_{ab \in \mathrm{imp}} \int_{-\infty}^{\infty}
G_{ab}(\iw) \Sigma_{ba}^\mathrm{dyn}(\iw) \D \omega \right]
\label{eq:e-total}
,
\end{align}
where $\Sigma^\mathrm{dyn}(\iw)$ is the dynamic part of the
lattice self-energy and $h$, $\gamma$, and $G(\iw)$, are the lattice
\hamiltonian{}, density-matrix, and Green's function.
The first term corresponds to the static Hartree--Fock-like contribution
and the second term takes dynamic self-energy contributions into account.
If the electron correlation does not lead to a breaking of the
occupation symmetry, i.e. the diagonal of the correlated
density matrix $\gamma_{aa}$ is constant, the first term becomes
identical to the Hartree--Fock energy and the second one becomes the
correlation energy of the method.
By calculating the energy in the lattice space,
instead of the cluster space, the explicit dependence of the
energy on the bath discretization is removed in both methods.
However, an implicit dependence on the bath discretization error remains
via the self-consistent density matrix, Green's function, and self-energy.

\section{Results} \label{sec:results}
We test the \wdmft{} method using the Hubbard model
on an infinite-dimensional Bethe~lattice, a 1D chain, and a 2D square lattice.
For all calculations, we use 800~Matsubara points and a fictitious temperature
of $T = 0.01$.
We note that this does not correspond to a finite temperature calculation,
but rather a choice of Matsubara discretization
since the impurity problem is being solved for the ground-state only.

\subsection{The Hubbard model on the infinite-dimensional Bethe lattice}\label{sec:bethe}
\begin{figure}
\includegraphics[width=\linewidth]{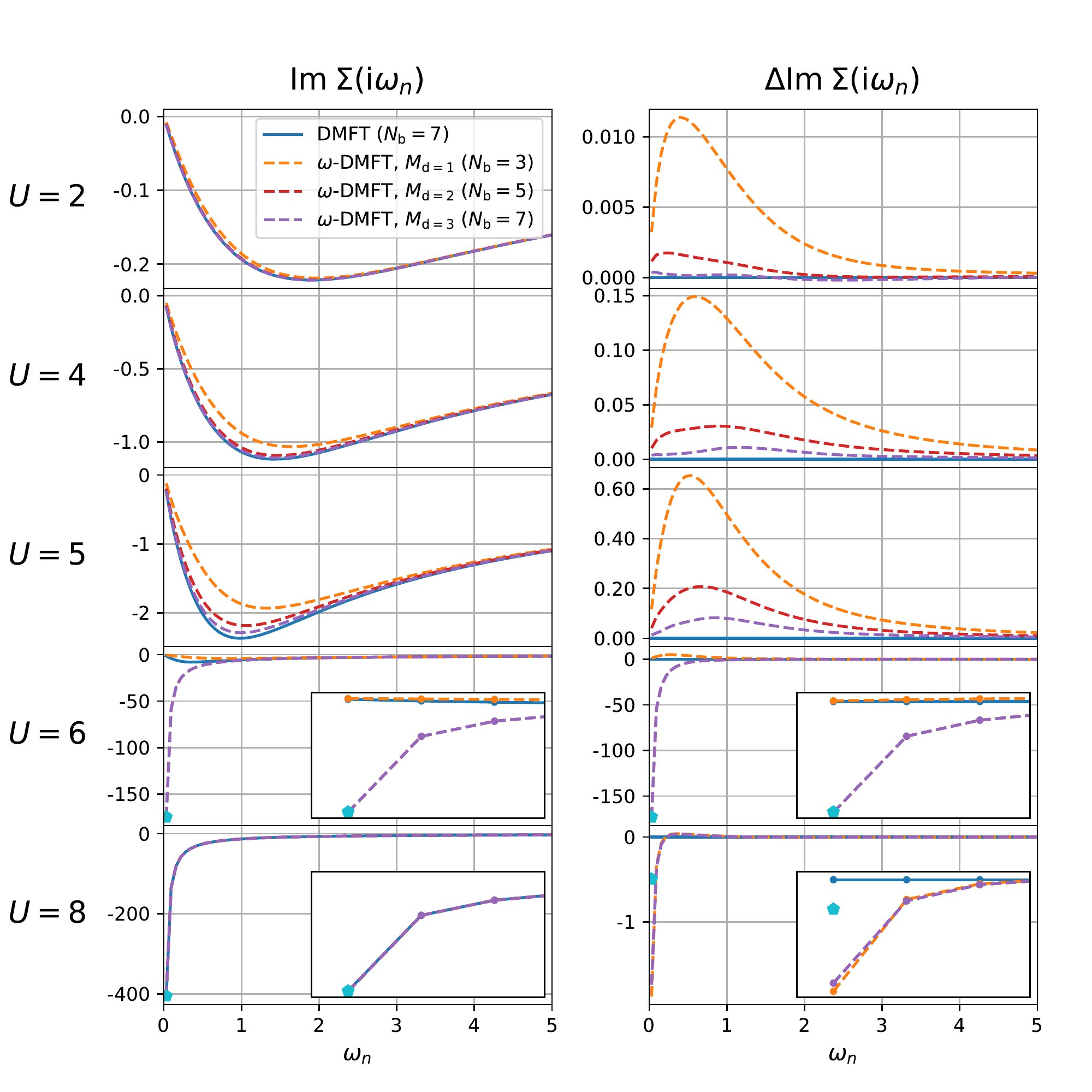}
\caption{
\label{fig:bethe-se}
Left column: Imaginary part of converged Matsubara self-energy for the
infinite-dimensional Bethe lattice Hubbard model.
The number in parenthesis indicates the number of bath orbitals.
Right column: Difference of the self-energy with respect to DMFT ($N_\mathrm{b}=7$).
For $U=6$ and $U=8$, the inset shows the first three Matsubara points to focus on
the low-energy regime, with the pentagon denoting the first Matsubara point of a DMFT calculation
with 9 bath orbitals to compare the convergence.
}
\end{figure}

\begin{figure}
\includegraphics[width=\linewidth]{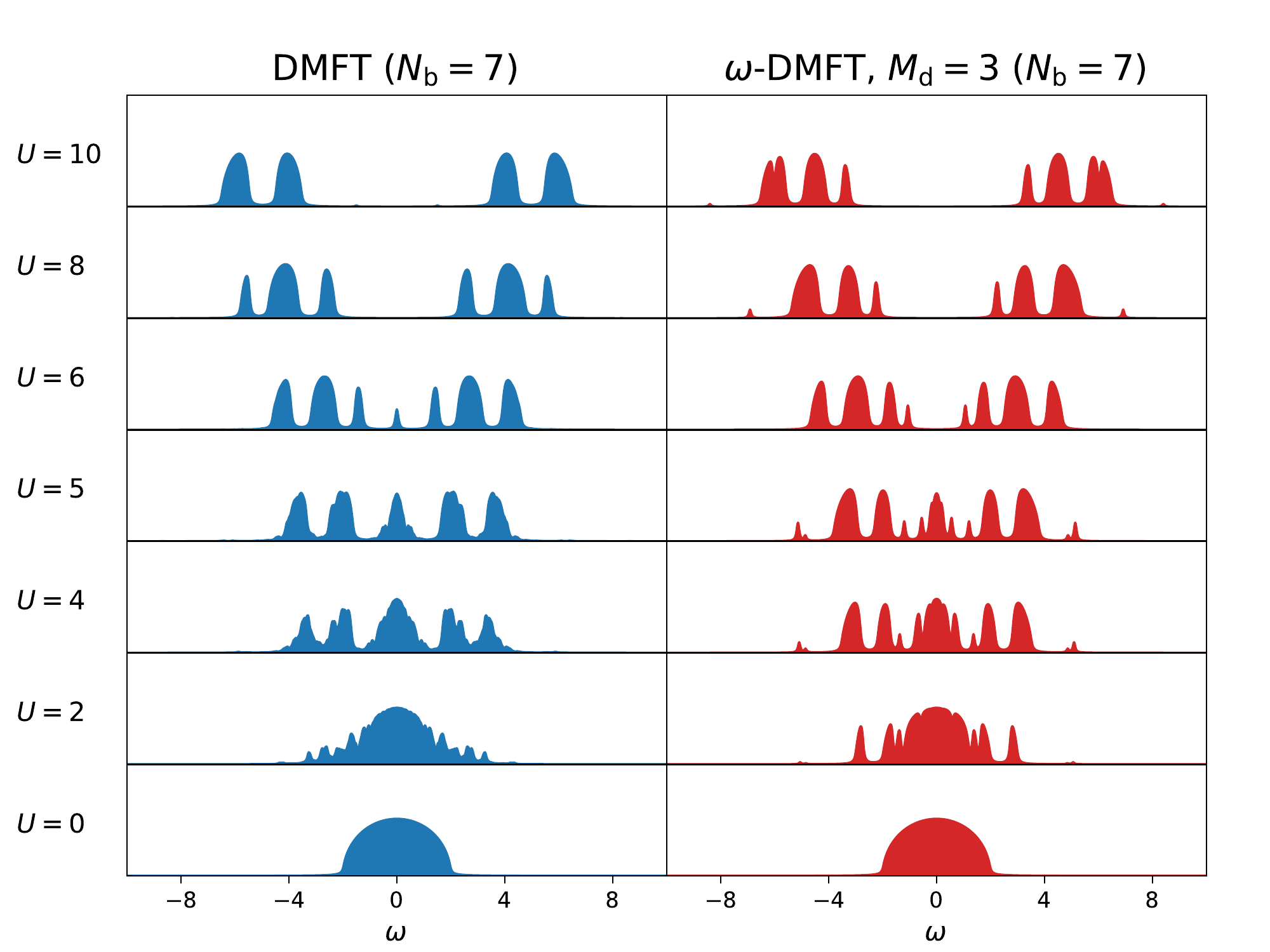}
\caption{
\label{fig:bethe-spectrum-aux}
Impurity DoS of the infinite-dimensional
Hubbard model on the Bethe lattice for DMFT with 7 bath orbitals,
and \wdmft{} with $M_\mathrm{d}=3$. While the transition to an insulating state is
made by $U=6$ within \wdmft{}, the DMFT solution is still erroneously metallic.
A Lorentzian broadening of $\eta = 0.05$ was used for both.
}
\end{figure}

The Hubbard model in infinite dimensions is an ideal benchmark
for quantum embedding methods, since the self-energy is purely
local and the DMFT approximation becomes exact.\cite{Georges1992}
In this model, it is found that at a critical value of $U$ there
exists a metal to Mott insulator quantum phase transition, with $U_C = 5.88$.
We model the Bethe lattice density of states (DoS) via 200 particle-hole symmetric states,
fitted to produce a semi-circular DoS, with every state being
subject to the Hubbard-$U$ interaction.\cite{KOLLAR2002}
In Fig.~\ref{fig:bethe-se} the convergence of the impurity self-energy
with respect to the dynamic bath order $M_\mathrm{d}$
is compared to a DMFT calculation with 7 bath orbitals
($M_\mathrm{d}=1$ has $3$ bath orbitals, $M_\mathrm{d}=2$ has $5$, and $M_\mathrm{d}=3$ has $7$),
showing both the converged self-energy, and the difference to DMFT with 7 bath orbitals.
The calculations were started from a self-energy converged at a slightly higher value of $U$,
when available.
At low~$U$ \wdmft{} self-energies are in excellent agreement with DMFT, even when a smaller number of
bath orbitals is used.
Just before the Mott-insulator-transition~(MIT) at $U = 5$ \
more dynamic bath orders are needed to recover the DMFT~self-energy.
At $U = 6$, DMFT with 7 bath orbitals and the \wdmft{} calculation with
$M_\mathrm{d} = 1$ favor the metallic solution favoured at lower $U$, whereas
\wdmft{} with $M_\mathrm{d} = 2$ and $M_\mathrm{d} = 3$ transition to the insulating solution.
The insulating solution at $U = 6$ is in agreement
with the critical value of $U_C = 5.88$ reported in
Ref.~\onlinecite{Bulla1999} using the numerical renormalization group method,
indicating that the \wdmft{} with $M_\mathrm{d} = 2$ and $3$ are more accurate than the corresponding DMFT
calculation with 7 bath orbitals. To corroborate this, we have also performed DMFT with
9 fitted bath orbitals, and find that the solution becomes
insulating with this enlarged bath space, demonstrating
the accuracy and compact nature of the bath space for \wdmft{}.
In both the insulating solutions at $U=6$ and $8$, the converged self-energy
for $M_\mathrm{d} = 2$ and $3$ are close to identical,
indicating that \wdmft{} is well converged with respect to the bath size, while
even at $U=8$, there still remain large differences seen in moving from 7 to 9 fit bath orbitals.
%
%
Overall, for this model, the \wdmft{} self-energy appears well converged
with $M_\mathrm{d} = 2$, resulting in 5 bath orbitals, whereas DMFT with 7 bath orbitals still has not
converged with respect to bath size, evidenced most clearly by the qualitatively incorrect solution at $U=6$,
as is apparent by comparison to DMFT with $N_\mathrm{b}=9$. This demonstrates the compact nature of the \wdmft{} bath space.
Figure~\ref{fig:bethe-spectrum-aux} shows the local DoS resulting
from the converged self-energy from
\wdmft{} using $M_\mathrm{d} = 3$ in comparison to DMFT with 7 bath orbitals.
The spectra are in broad agreement for all values of $U$, while
at $U = 6$ the DMFT solution still has a small quasiparticle peak,
whereas the \wdmft{} solution is fully insulating.
The \wdmft{} spectral functions can also show more high-energy features than DMFT,
although it is not clear if these are physical, or a result of the auxiliary
representation of the self-energy.

\subsection{The 1D Hubbard model}\label{sec:hub-1d}
\begin{figure}
\includegraphics[width=\linewidth]{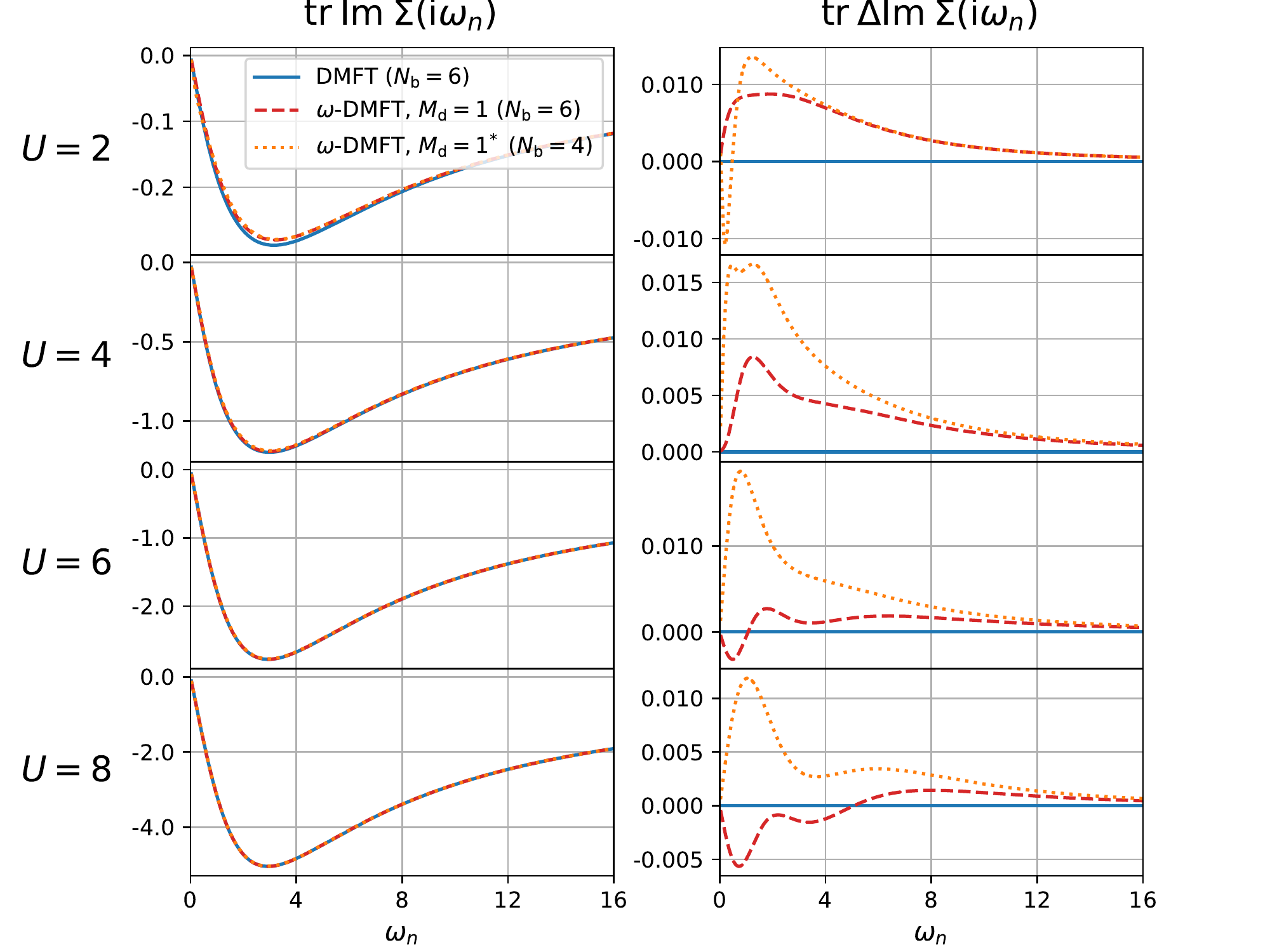}
\caption{
\label{fig:hub1d-s-imp}
Similar to Fig.~\ref{fig:bethe-se} but for the 1D Hubbard model with a
2-site impurity. The trace of the self-energy is shown. Discrepancies between DMFT
and \wdmft{} are exceptionally small across all correlation regimes, even utilizing the
additional orbital factorization approximation.
}
\end{figure}

\begin{figure}
\includegraphics[width=\linewidth]{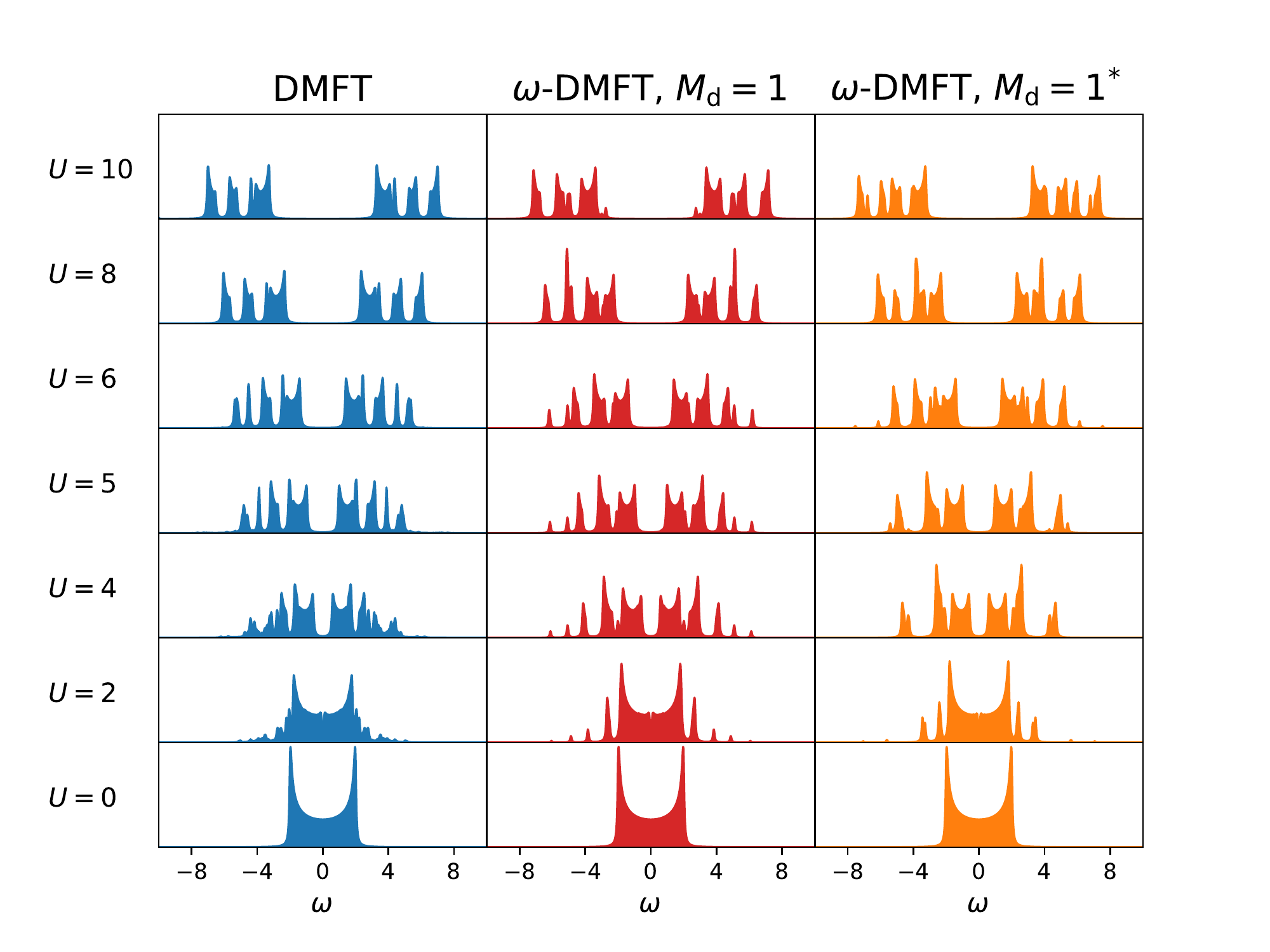}
\caption{
\label{fig:hub1d-spectrum-aux}
Trace of the local DoS of the 1D Hubbard model with a 2-site impurity.
A broadening of $\eta = 0.05$ was used.
}
\end{figure}

\begin{figure}
\includegraphics[width=\linewidth]{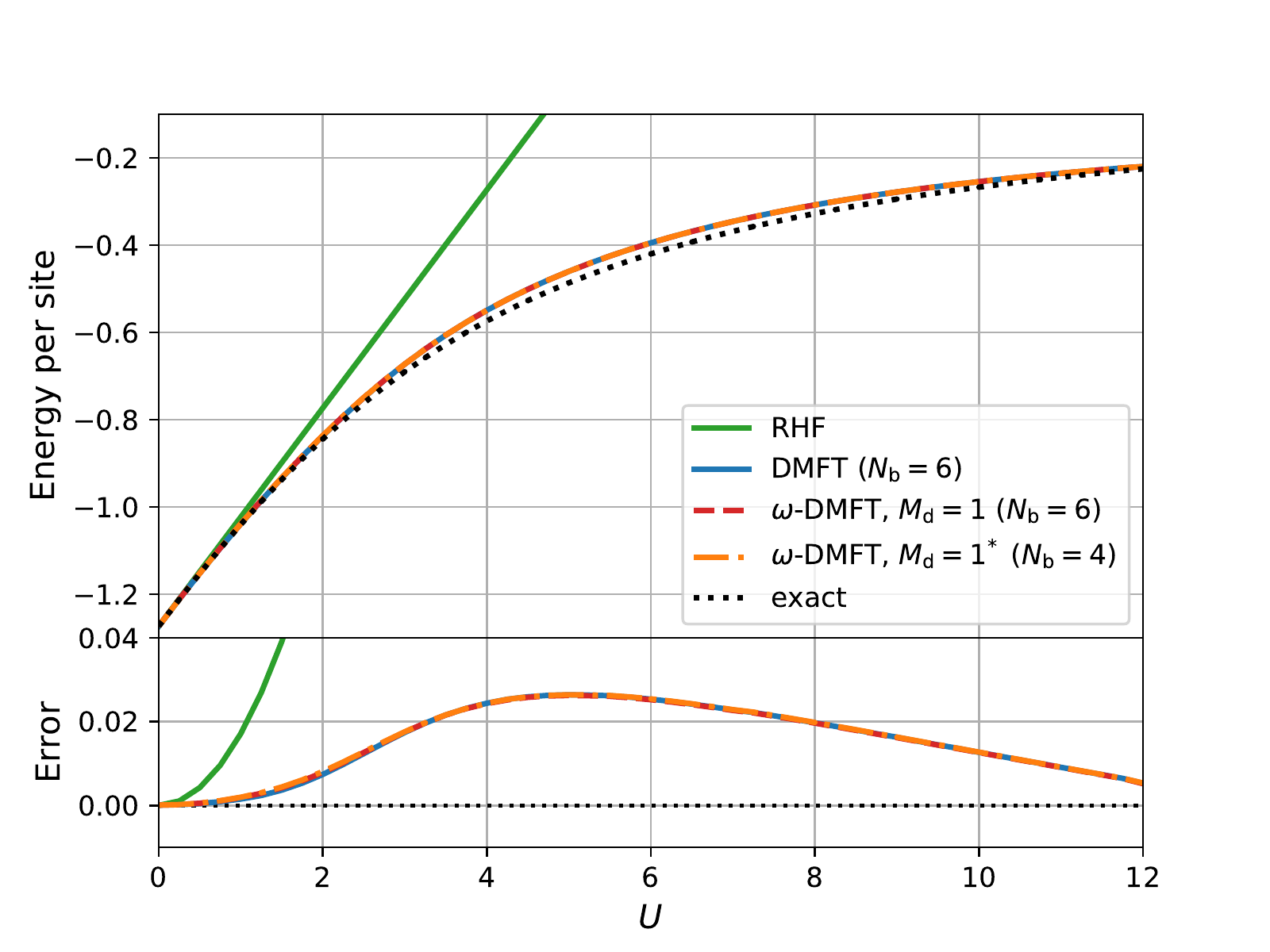}
\caption{
\label{fig:hub1d-energy}
Top row: Ground-state energy per site of the 1D Hubbard~model, calculated according to
Eq.~\eqref{eq:e-total}.
Bottom row: Error in the total energy, compared to the exact solution from
the Bethe ansatz.
}
\end{figure}

The 1D Hubbard chain is a difficult system for single-site DMFT,
since both short and long-range correlation are important, and the DMFT
approximation is far from its exact infinite-dimensional limit.
Here we consider an impurity cluster of 2 sites, so that nearest neighbor
correlations can be captured explicitly within the impurity model.
This cluster also allows for assessment of the accuracy of the impurity-orbital
factorization of the bath space.
The full lattice consists of 400~sites with anti-periodic boundary conditions,
so that finite-size effects of the lattice become negligible.
Figure~\ref{fig:hub1d-s-imp} compares the trace of the Matsubara
self-energies obtained with DMFT ($N_\mathrm{b} = 6$) and \wdmft{} using $M_\mathrm{d} = 1$,
which results in 2 static and 4 dynamic bath orbitals.
Additionally, we also test the orbital factorization
(denoted $M_\mathrm{d} = 1^*$) of the dynamic bath space
according to \eq{}\eqref{eq:wdmft-bath-orbfac},
which leads to only 2 static and 2 dynamic bath orbitals.
Evidently, the 1D Hubbard model requires a smaller number of
bath orbitals for a good description of the hybridization than the
(formally) infinitely coordinated Bethe lattice, which can be argued on
simple quantum information grounds.
The self-energies of all three methods are very similar, with only very minor
discrepancies between the methods at all $U$, where it is unclear which method is more
accurate.
%
The orbital factorization additionally only leads to very small changes in the self-energy.
Note that although Fig.~\ref{fig:hub1d-s-imp} only shows the trace
of the self-energy, it will still be affected by any errors in the
off-diagonal parts of the impurity Green's~function
resulting from the orbital factorization,
due to the inversion of the Green's~function in the Dyson~equation.
In Figure~\ref{fig:hub1d-spectrum-aux}, the spectral functions
of the three different methods are shown.
Again, DMFT and \wdmft{} spectral functions are almost identical,
especially in the low frequency regime.
Finally, in Fig.~\ref{fig:hub1d-energy}, we compare the total energy per site
of DMFT and \wdmft{} to the exact thermodynamic limit energy obtained from the
Bethe ansatz.\cite{Lieb1968}
To illustrate the strength of the electron correlation at different values of $U$,
we also show the restricted Hartree--Fock (RHF) energy.
All energy curves lie on top of each other and are very close to
the exact energy at low and high values of $U$.
In the intermediate region, both DMFT and \wdmft{}
underestimate the total energy, due to the neglect of long-range correlation.
In conclusion, both the standard \wdmft{} and the orbital-factorized variation
agree with the results of DMFT for the two-impurity 1D Hubbard model without appreciable error.

\subsection{The 2D Hubbard model}\label{sec:hub-2d}
The 2D Hubbard model on a square lattice is a challenging system for embedding methods.
On the one hand, the dimensionality is not high enough to make single-site DMFT
accurate, as it is in the case of the Bethe lattice.
On the other hand, more bath states are needed to describe the hybridization
than in the one-dimensional case, severely restricting either the impurity size or the bath size
of cluster DMFT calculations.
Although the true ground-state is expected to be gapped at any finite-$U$ value, for finite clusters
embedded in a paramagnetic phase, the absence of fixed (static) magnetic moments developing in the environment constrains the description of the long-range
spin fluctuations, and precludes the opening of a gap at low $U$\cite{Schafer2015}. Therefore, when restricted to paramagnetic phases, the model 
develops a finite-$U$ phase transition from metallic to Mott-insulating state, which is used as
an example of the physical processes governing this correlation-driven behavior present in many 
two-dimensional systems. This universal behavior is also seen in the restriction of QMC wavefunction 
ansatz to paramagnetic forms.\cite{Gros11}

\begin{figure}
\includegraphics[width=\linewidth]{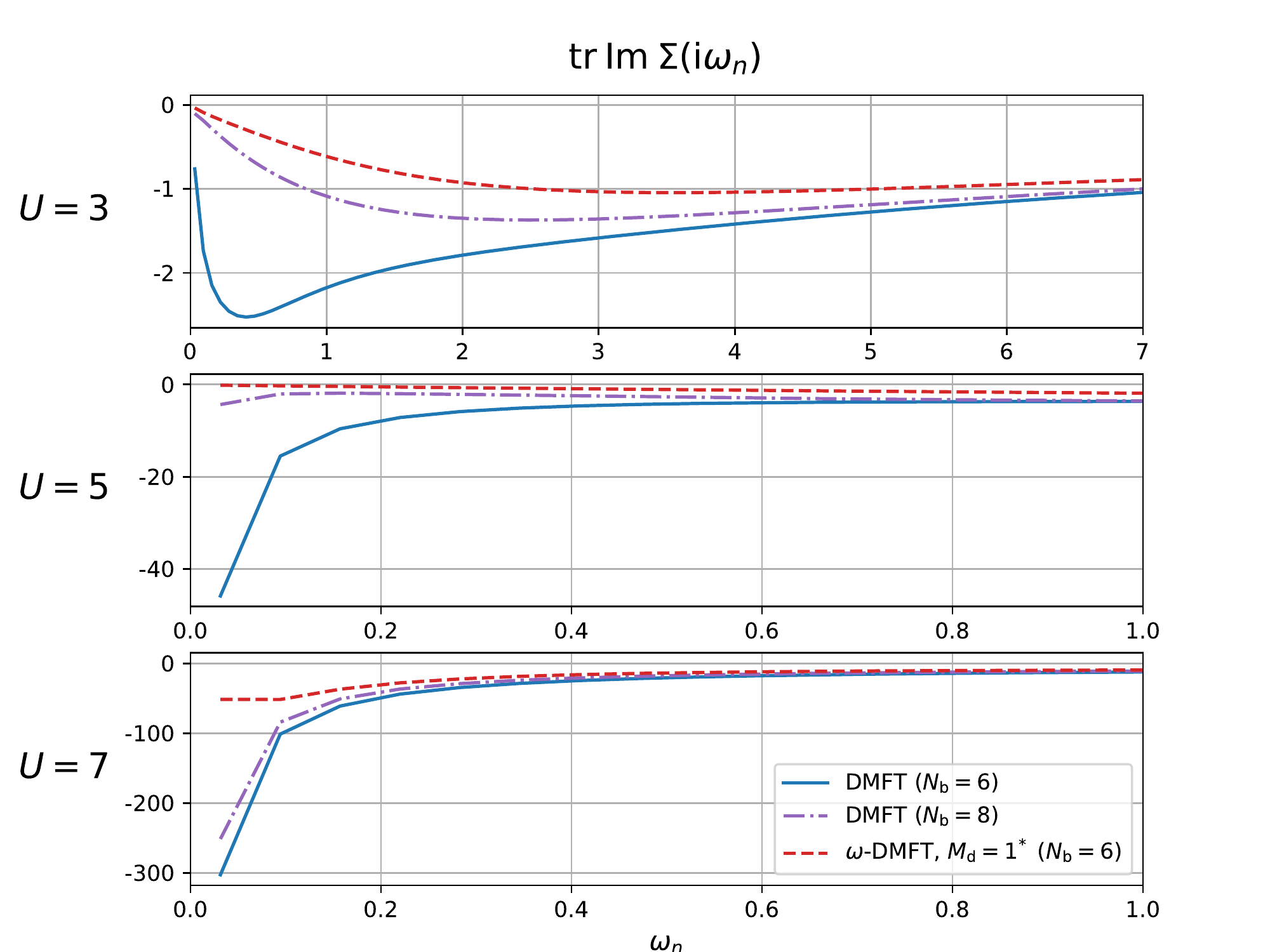}
\caption{
\label{fig:hub2d-s-imp}
Trace of the imaginary part of the impurity Matsubara self-energy of the 2D Hubbard lattice for three representative correlation strengths,
for DMFT with $N_\mathrm{b}=6$ and 8 orbitals, and \wdmft{} with 6 bath orbitals.
Note the different scales of the frequency axes, allowing a clearer resolution of the differences in the low energy regime for higher $U$.
}
\end{figure}

\begin{figure}
\includegraphics[width=\linewidth]{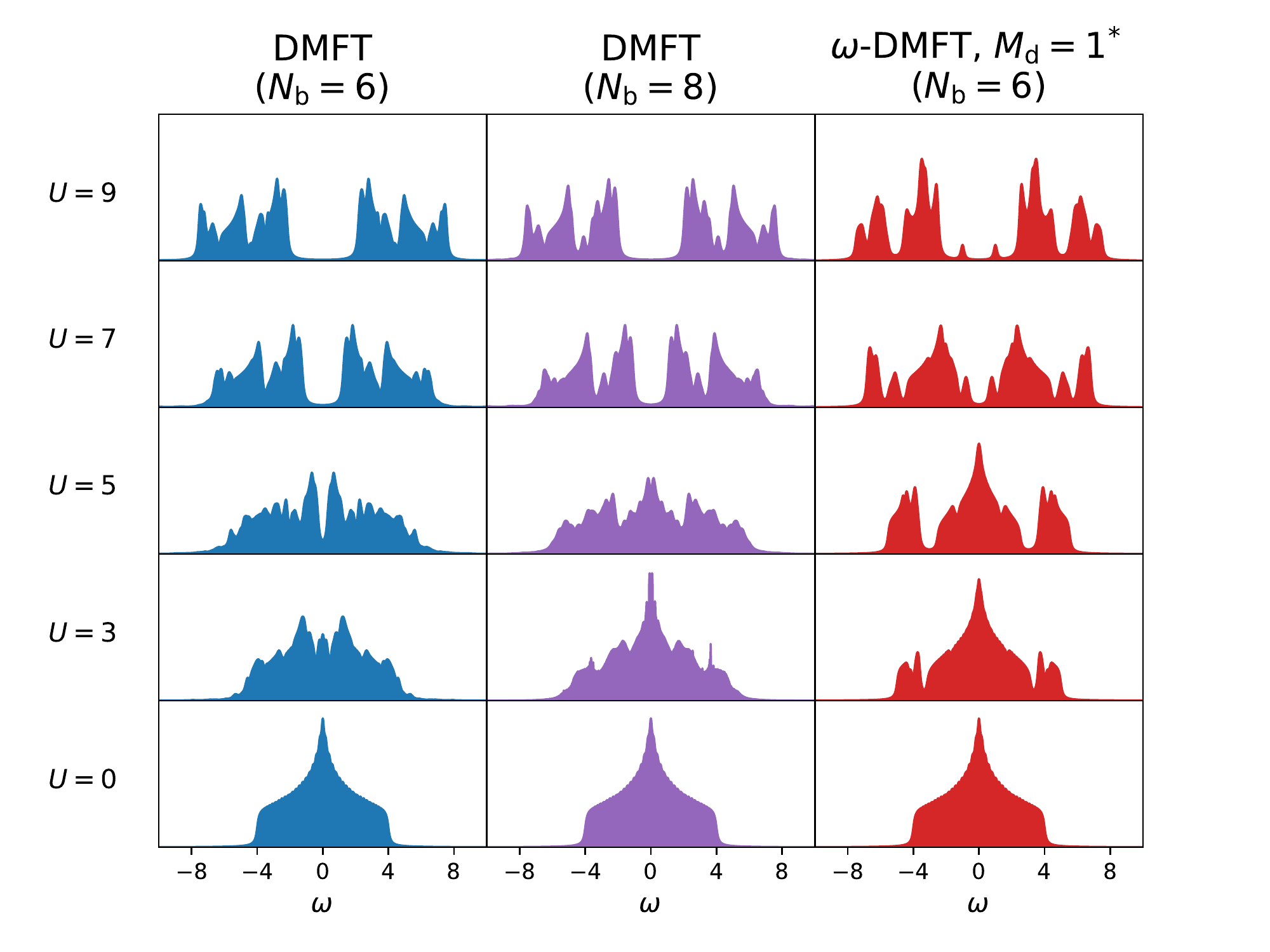}
\caption{
\label{fig:hub2d-spectrum-aux}
Trace of the impurity-projected lattice DoS for the 2D Hubbard model.
The calculations were converged on a $48 \times 24$ lattice and the resulting
self-energy (DMFT) or auxiliaries (\wdmft{}) were put on a $60 \times 60$ lattice,
to obtain smoother spectra.
A broadening of $\eta = 0.1$ was used.
}
\end{figure}

Here we consider square fragments of $2 \times 2$ sites on a lattice
of size $48 \times 24$ with anti-periodic boundary conditions in both dimensions.
An additional difficulty in standard ED-DMFT calculations is that the lattice geometry
can put constraints on the bath parameterization.
These can be taken into account, but this requires knowledge about the system at hand
and reduces the functional space onto which the hybridization is projected.
We chose to not apply any symmetry constraints to the bath parameters,
except for those due to particle-hole symmetry as mentioned in Sec.~\ref{sec:hub-1d}.
For the \wdmft{} method, we applied the orbital factorization
(denoted $M_\mathrm{d} = 1^{\!*}$), resulting in 6 bath orbitals.
Figure~\ref{fig:hub2d-s-imp} compares the trace of the imaginary part of the
impurity self-energy of DMFT with 6 bath orbitals, 8 bath orbitals, and \wdmft{} (6 bath orbitals).
At $U = 3$, \wdmft{} underestimates the self-energy in comparison to
DMFT~($N_\mathrm{b} = 8$), but yields
a self-energy that is closer to DMFT~($N_\mathrm{b} = 8$)
than a DMFT calculation with only 6 bath orbitals, which substantially
overestimates the correlation.
This manifests in the spectrum (Fig.~\ref{fig:hub2d-spectrum-aux})
as an erroneous removal of the sharp quasiparticle peak derived from the van Hove singularity
at $\omega=0$ for DMFT ($N_\mathrm{b}=6$), which is retained in both
the 8~bath orbital results, and the \wdmft{} approach demonstrating the
compactness of this bath space.
The behavior is exacerbated at $U = 5$, where DMFT with 6~bath orbitals clearly has an insulating solution, with distinct spectral gap and a divergent self-energy.
Increasing this to 8~bath orbitals changes the character to mostly
metallic, except for the first Matsubara point which indicates a potential divergence, but a spectrum which looks metallic.
In contrast, the \wdmft{} self-energy is entirely metallic and much closer to the
DMFT~($N_\mathrm{b} = 8$) result than DMFT ($N_\mathrm{b} = 6$).
The spectrum of Fig.~\ref{fig:hub2d-spectrum-aux} at this $U$ value
retains the sharp feature at the Fermi level, with a similar height as the ones for lower values of $U$ (and thus fulfilling the Luttinger theorem). The 
higher frequency part of the self-energy then serves to develop significant Hubbard side-bands in the spectrum. This character is anticipated and 
follows the trend of increasing numbers of bath orbitals in DMFT, as well as the comparable literature DMFT suggesting a metallic solution at this $U$\cite{Park2008}.
At $U = 7$ and $U = 9$, \wdmft{} again has a smaller value of the self-energy in the low-frequency regime
than both DMFT self-energies, resulting in a slightly smaller gap, with additional small low-energy features. 
However all approaches result in a gapped solution.
We note, however, that at all values of $U$ the DMFT self-energy becomes more similar
to the \wdmft{} self-energy as the number of bath orbitals is increased.
%
%
%
%
%

\begin{figure}
\includegraphics[width=\linewidth]{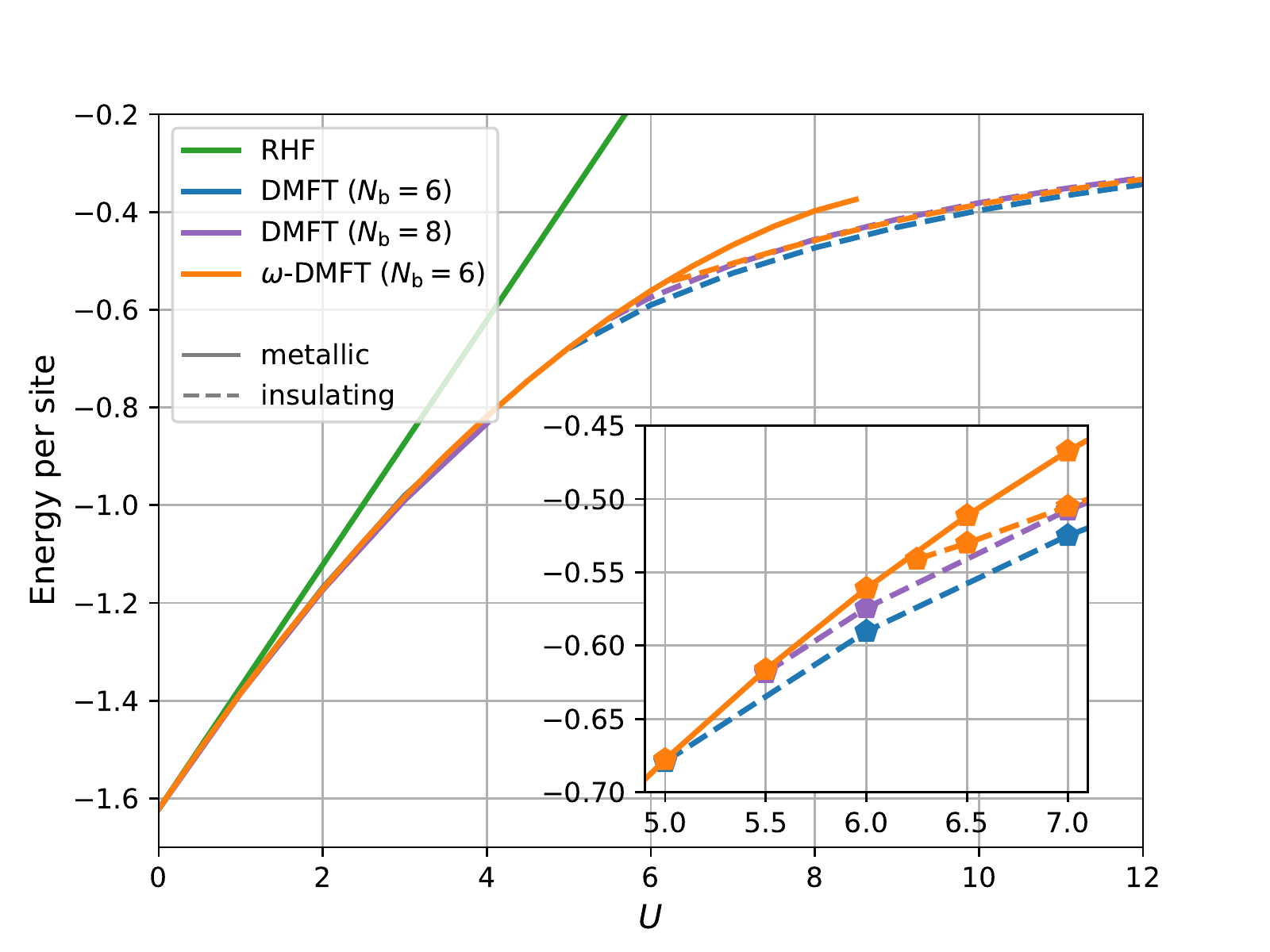}
\caption{
\label{fig:hub2d-energy}
Energy per site of the 2D Hubbard model, calculated according to Eq.~\eqref{eq:e-total}, for
DMFT with $N_\mathrm{b}=6$ and 8 orbitals, and \wdmft{} with 6 bath orbitals. Metallic solutions are given by
solid lines, while the insulating solutions are dashed. The inset shows a zoom of the transition region, showing
the slightly higher $U$ metal-insulator transition for the \wdmft{} than DMFT, although the larger number of bath orbitals
tends towards this value.
}
\end{figure}

%
Finally, \fig{}\ref{fig:hub2d-energy} shows the total~energy per site.
In the metallic regime below the Mott transition, the energies of DMFT and \wdmft{} are almost
identical.
Above the transition, the \wdmft{} energy is close to the DMFT curve obtained with
8~bath orbitals, while DMFT with 6~bath orbitals underestimates the energy by
approximately $0.015t$.
The MIT occurs between $U = 6$ and $U = 6.25$ in \wdmft{} and no metastable insulating phase
was found for $U < U_C$.
On the other hand, the metallic \wdmft{} solution is metastable up to $U_{C_2} = 8.5$.
The slope of the total energies changes discontinuously between metallic and insulating
solution at $U_C$ (see inset of \fig{}\ref{fig:hub2d-energy}), which indicates
a first-order phase transition.
In our DMFT calculation with 8 bath orbitals,
the insulating solutions is stable down to $U = 5.5$ and no discontinuity of the derivative
of the energy is visible.
The non-existence of a metastable insulating phase below $U_C$ is in agreement
with the cluster DMFT results from Ref.~\onlinecite{Park2008},
which also suggest that the zero-temperature transition $U_C$
coincides with the lower second order critical point $U_{C_1}$.
However, the values of the critical points reported in Ref.~\onlinecite{Park2008} are lower
($U_{C_1} \approx 5.8$ and $U_{C_2} \approx 6.25$). These discrepancies could be due to the
finite-temperature extrapolation of this work, or the remaining bath discretization error in the
\wdmft{}.
%


\section{Conclusion} \label{sec:conclusion}
We have shown a way to algebraically construct frequency-dependent
bath orbitals, which are designed to exactly reproduce the hybridization
of the impurity to its environment, and be systematically expandable in the presence of interactions.
These dynamic bath orbitals can be used in the self-consistent \wdmft{} method presented here,
avoiding the difficult non-linear fit required in most ED-DMFT formulations.
We show that \wdmft{} can reproduce the self-energies, density-of-states,
and total energies of DMFT calculations for the Hubbard model on
the infinitely coordinated Bethe lattice and the two-impurity 1D chain across all correlation strengths.
In the case of the Bethe~lattice, \wdmft{} correctly places the
Mott-Insulator transition between $U = 5$ and $U = 6$
with only 5~bath~orbitals.
On the other hand, in standard ED-DMFT 9~bath~orbitals are
needed to place the transition in this interval, with fewer
bath~orbitals resulting in too high a transition between
$U = 6$ and $U = 7$.
We also applied \wdmft{} to the 2D square lattice with a $2\times2$ impurity cluster,
employing an orbital factorization to obtain a manageable bath size of only 6 orbitals.
In the metallic regime the self-energies of \wdmft{} were shown
to be in better agreement with DMFT results with 8~bath orbitals, than
a DMFT calculation with 6~bath orbitals.
In the insulating regime, \wdmft{} self-energies and density-of-states
differ from the DMFT results at low frequency, although the latter
tend towards \wdmft{} with increasing bath size.
The total energy of \wdmft{} is in good agreement with DMFT with 8 bath orbitals, with
the first-order phase transition predicted to be between $U = 6$ and $U = 6.25$
and thus slightly higher than in DMFT with a finite-temperature CT-QMC solver ($U_C \approx 5.8$ in Ref.~\onlinecite{Park2008}).

The weight of numerical evidence across these different systems indicates
that the frequency-dependent bath of \wdmft{}
can allow for a more compact representation of the hybridization, allowing for a more faithful
representation of the impurity problem, leading to excellent results, with lower computational resources.
In the future, we hope to combine this approach with emerging approximate Hamiltonian
impurity solvers, which will allow the application to larger impurity spaces, and a rigorous expansion of the dynamic
bath space in \wdmft{}. This will allow application to systems with long-range interactions, including moving towards {\em ab initio} correlated materials.
Furthermore, for solvers which would not work well computing the Green's function at individual frequency points (e.g. imaginary-time solvers), the
method can be recast to analytically compute {\em time-dependent} bath orbitals in a systematic fashion, which will be investigated in the future.

\section{Acknowledgements}

G.H.B. gratefully acknowledges support from the Royal Society via a University Research Fellowship, in addition to funding from the European Union's Horizon 2020 research and innovation programme under grant agreement No. 759063. We are also grateful to the UK Materials and Molecular Modelling Hub for computational resources, which is partially funded by EPSRC (EP/P020194/1).

\end{document}